\documentclass[12pt,doublespacing]{article}

\usepackage{fullpage}
\usepackage{setspace}
\usepackage{authblk}

\usepackage{cite}
\usepackage{hyperref} 

\usepackage{graphicx}

\usepackage{amsmath}
\usepackage{amssymb}
\allowdisplaybreaks

\usepackage{float}
\usepackage{subfig}
\usepackage{makecell}
\usepackage{rotating}

\usepackage{enumitem}

\begin{document}

\title{Characteristic Mode Analysis of Composite Nanostructures using a Coupled System of Volume Integral and Hydrodynamic Equations}

\author[1]{Meruyert Khamitova}
\author[2]{Ran Zhao}
\author[3]{Doolos Aibek Uulu}
\author[1]{Sebastian Celis Sierra}
\author[1]{Hakan Bagci\vspace{0.5cm}}

\affil[1] {Meruyert Khamitova, Sebastian Celis Sierra, and Hakan Bagci are with the Electrical and Computer Engineering Program, Computer, Electrical, and Mathematical Science and Engineering (CEMSE) Division, King Abdullah University of Science and Technology (KAUST), Thuwal 23955-6900, Saudi Arabia (e-mail: 
meruyert.khamitova@kaust.edu.sa;
sebastian.celissierra@kaust.edu.sa;  hakan.bagci@kaust.edu.sa).\vspace{0.5cm}}

\affil[2]{Ran Zhao was with the Electrical and Computer Engineering (ECE) Program, Computer, Electrical, and Mathematical Science and Engineering (CEMSE) Division, King Abdullah University of Science and Technology (KAUST), Thuwal 23955-6900, Saudi Arabia. He is now with the School of Electronic Science and Engineering, University of Electronic Science and Technology of China (UESTC), Chengdu 61173, China (e-mail: ran.zhao@uestc.edu.cn).\vspace{0.5cm}}

\affil[3]{Doolos Aibek Uulu was with the Electrical and Computer Engineering (ECE) Program, Computer, Electrical, and Mathematical Science and Engineering (CEMSE) Division, King Abdullah University of Science and Technology (KAUST), Thuwal 23955-6900, Saudi Arabia. He is now with the Department of Cyber Security, Light Academy College of Engineering, Bishkek, Kyrgyzstan (e-mail: doolosaibekuulu@gmail.com).}

\date{}
\maketitle
\newpage

\begin{abstract}
Full-structure and sub-structure characteristic mode analysis (CMA) formulations are developed for composite metallic--dielectric nanostructures based on a coupled system of volume integral equations (VIE) and the hydrodynamic equation (HDE). In the full-structure CMA, the generalized eigenvalue equation (GEE) is constructed from the matrix of the complete coupled system, and the resulting characteristic currents describe the response of the entire composite nanostructure. In the sub-structure CMA, the GEE is constructed from a reduced system, derived from the coupled system by eliminating the dielectric-region unknowns, so that it is expressed only in terms of the metallic-region currents. This isolates the resonances of the metallic region while still incorporating the effect of the dielectric region through the reduced system. Because the reduced system has a smaller dimension, the sub-structure CMA is computationally more efficient than the full-structure CMA and, when the dielectric does not resonate in the frequency range of interest, identifies the same resonances. Both formulations are validated against extinction cross-section (ECS) results and are used to characterize how a dielectric environment reshapes the metallic resonances, including substrate-induced red-shifts and, for high-contrast substrates in direct contact, hybridized modal responses.

\end{abstract}

\newpage

\section{Introduction}
Metallic nanostructures support localized surface plasmon resonances that confine electromagnetic fields to subwavelength regions, producing strong near-surface field enhancement. This property has made them building blocks of photonic devices for applications such as biosensing, bioimaging, and solar energy harvesting~\cite{stockman2015nanoplasmonic, zhang2013enhancement,Novotny2011,west2003engineered}. Although optical absorption and heat generation in metallic nanostructures are useful for thermoplasmonic applications and medical therapies~\cite{Baffou2013ThermoPlasmonics,govorov2007generating}, they can also reduce the efficiency of plasmonic devices in applications where low loss is desired~\cite{barreda2022applications,barreda2021metal}. In contrast, high-refractive-index dielectric nanostructures exhibit negligible absorption, allowing light to propagate with significantly lower energy dissipation~\cite{barreda2022applications,krasnok2012all,krasnok2018spectroscopy}, but provide weaker near-field enhancement than metallic nanostructures. In addition, the enhanced fields in dielectric nanoparticles are predominantly confined within the particles, which can limit their effectiveness in sensing applications~\cite{lepeshov2019hybrid}. Composite metallic--dielectric nanostructures can combine the advantages of both materials: the strong near-field enhancement of metals and the low-loss scattering response of dielectrics~\cite{lepeshov2019hybrid}. Such composite structures provide an effective approach for tuning and optimizing the resonance behavior of metallic nanostructures. For example, the presence of a dielectric substrate can red-shift metallic resonance modes, with the magnitude of the shift depending on the substrate permittivity and its distance from the metallic structure~\cite{sherry2005localized}. For high-permittivity substrates or large contact areas between metallic and dielectric regions, the original metallic resonances can be significantly modified, giving rise to substrate-mediated hybridization of modes~\cite{zhang2011substrate,sherry2005localized}.

When a composite nanostructure is under electromagnetic excitation, free-electron current is induced in the metallic region~\cite{forstmann1986metal}, while bound-electron polarization current is induced in the dielectric region. Near the plasma frequency, the response of free electrons in the metallic region becomes nonlocal, leading to spatial dispersion~\cite{mortensen2013}. This nonlocality is not captured by the local Drude model~\cite{mortensen2012}, and thus the hydrodynamic equation (HDE) is commonly used as an extended framework to describe the free-electron response~\cite{raza2015nonlocal}. In contrast, the bound-electron response in the dielectric region is usually treated as local because the electrons are tightly bound to the atomic nuclei~\cite{zheng2018boundary}. The free-electron current, also termed the hydrodynamic current, and the bound-electron current in the two regions are coupled to the electromagnetic fields via the Maxwell equations~\cite{forstmann1986metal}. The electromagnetic response of the composite metallic--dielectric nanostructures can therefore be described by incorporating the HDE into the Maxwell equations. Recently, a coupled system of the volume integral equation (VIE) and the HDE has been formulated to analyze scattering from such structures~\cite{uulu2023coupled}. This approach allows the resonance behavior of the composite nanostructure to be analyzed under a specific excitation. However, because the response depends on the excitation, it does not directly provide an excitation-independent modal description of all resonances supported by the structure.

Characteristic mode analysis (CMA) is a reliable tool for analyzing the intrinsic resonant modes of arbitrarily shaped objects. A key advantage of CMA is that it provides a physical interpretation of characteristic eigenvalues and their corresponding characteristic modes in terms of the radiated, dissipated, and reactive powers~\cite{chen2015characteristic, ylaoijala2019generalized}. Although CMA was initially developed for conducting bodies~\cite{harrington1971theory, harrington1971computation}, it has been successfully extended to dielectric, magnetic, and composite metallic--dielectric structures~\cite{harrington1972characteristic, guo2018generalized, huang2021accurate}. In CMA, a generalized eigenvalue equation (GEE), derived from the matrix system obtained by discretizing the governing integral equation, is solved at each discrete frequency point over the frequency range of interest. This provides an excitation-independent modal description and enables the identification of resonances supported by the structure.

More recently, CMA has been extended to isolated metallic nanostructures~\cite{khamitova2024characteristic, khamitova2026characteristic}. Because an isolated metal supports only the free-electron current, a single hydrodynamic volume integral equation (HDVIE) in terms of the induced current can be obtained by inserting the VIE into the HDE. The HDVIE is then cast as a GEE. This reduction is no longer possible for composite metallic--dielectric structures, where the bound-electron current in the dielectric region introduces a second, coupled unknown. In such cases, the coupled VIE--HDE system must be retained.

In this work, full-structure and sub-structure CMA formulations are developed using the coupled VIE--HDE system to investigate the resonance behavior of composite metallic--dielectric nanostructures. Applying CMA directly to the coupled system yields a full-structure CMA formulation, in which the GEE is constructed from the matrix of the discretized coupled system for the complete composite structure. The resulting characteristic modes describe the response of the entire structure. However, in this formulation, it is not straightforward to isolate whether a given mode is primarily associated with the metallic region or the dielectric region~\cite{huang2021accurate}. To address this issue, a sub-structure CMA formulation is developed, in which the dielectric-region unknowns are eliminated from the coupled system and a reduced matrix is constructed for the metallic region. The resulting GEE is expressed solely in terms of the metallic-region current unknowns, while the effect of the dielectric environment is retained through the reduced matrix. This approach isolates the resonances associated with the metallic nanostructure. Moreover, because the reduced GEE has a smaller dimension than the full-structure GEE, the sub-structure CMA can reduce the computational cost of the eigensolution. However, this advantage depends on the cost of constructing the reduced matrix, which can become significant for composite structures with large dielectric regions. A preliminary version of this work is reported in~\cite{khamitova2025characteristic}.

A notable property of CMA is that it provides real, or equiphase, characteristic currents, which generate equiphase electric fields with a constant phase lag~\cite{chen2015characteristic}. This behavior is important in scattering problems because it allows the characteristic modes to convert incoming waves into outgoing waves through modal amplitude and phase changes~\cite{huang2021full}. In several sub-structure CMA formulations, the modes associated with the sub-structure are equiphase, whereas the modes of the complete structure are not necessarily in phase~\cite{huang2021accurate, alroughani2014classification, alroughani2016orthogonality}. In this work, both full-structure and sub-structure CMA formulations are developed to ensure the real, or equiphase, nature of the characteristic currents. This enables a consistent comparison between the modes obtained from the two formulations.

The remainder of this paper is organized as follows. Section~\ref{sec:formulation} presents the coupled VIE--HDE system, its discretization, and the proposed full-structure and sub-structure CMA formulations. Section~\ref{sec:num_results} provides numerical examples that validate the proposed formulations and examine the influence of dielectric media on the resonance behavior of metallic nanostructures. Finally, the conclusions are drawn in Section~\ref{sec:conclusion}.

\section{Formulation}
\label{sec:formulation}

\subsection{Coupled VIE--HDE System}
\begin{figure}[ht!]
\centering
  \includegraphics[width=0.4\columnwidth]{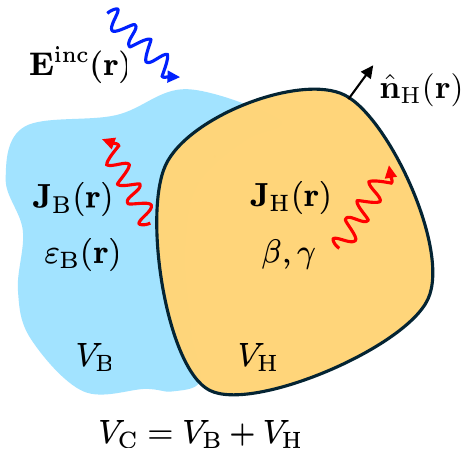}
  \caption{Description of the electromagnetic problem where $V_\mathrm{C}$ represents the volume of the arbitrarily-shaped composite nanostructure.}\label{fig:problem_description}
\end{figure}

Let $V_\mathrm{C}$ denote a composite nanostructure that consists of a dielectric region $V_{\mathrm{B}}$ and a metallic region $V_{\mathrm{H}}$, such that $V_{\mathrm{C}}=V_{\mathrm{B}} \cup V_{\mathrm{H}}$, as depicted in Fig.~\ref{fig:problem_description}. The composite nanostructure resides in a homogeneous, unbounded background medium with permittivity $\varepsilon_0$ and permeability $\mu_0$. The dielectric region $V_{\mathrm{B}}$ contains only bound electrons, while the metallic region $V_{\mathrm{H}}$ is assumed to contain only free electrons.

The composite nanostructure $V_{\mathrm{C}}$ is excited by an incident electric field $\mathbf{E}^\mathrm{inc}(\mathbf{r})$. An $e^{\mathrm{j}\omega t}$ time-harmonic dependence is assumed and suppressed throughout, where $\omega$ is the angular frequency. In response to this excitation, equivalent volumetric electric current densities $\mathbf{J}_{\mathrm{B}}(\mathbf{r})$ and $\mathbf{J}_{\mathrm{H}}(\mathbf{r})$ are induced within $V_\mathrm{B}$ and $V_\mathrm{H}$, respectively~\cite{chew1999waves,jin2015}. These current densities generate the scattered electric field $\mathbf{E}^\mathrm{sca}(\mathbf{r})$ as 
\begin{equation}
    \mathbf{E}^{\mathrm{sca}}(\mathbf{r}) =\mathcal{L}_{V_\mathrm{B}}[ \mathbf{J}_{\mathrm{B}}](\mathbf{r})+\mathcal{L}_{V_\mathrm{H}}[ \mathbf{J}_{\mathrm{H}}](\mathbf{r}).
    \label{eq:e_sca}
\end{equation}
Here, the integral operator $\mathcal{L}_{V}[\mathbf{X}](\mathbf{r})$ is given by
\begin{equation}
\label{eq:l_operator}
\begin{aligned}
\mathcal{L}_{V}[\mathbf{X}](\mathbf{r})= &-\mathrm{j} \omega \mu_0 \int_V \mathbf{X}(\mathbf{r}^{\prime}) g_0(\mathbf{r}, \mathbf{r}^{\prime})\,dv^{\prime}+\frac{1}{\mathrm{j} \omega \varepsilon_0} \nabla\int_V \nabla^\prime \cdot \mathbf{X}(\mathbf{r}^{\prime}) g_0(\mathbf{r}, \mathbf{r}^{\prime})\,dv^{\prime}\\&-\frac{1}{\mathrm{j} \omega \varepsilon_0} \nabla\int_{\partial V} \hat{\mathbf{n}}(\mathbf{r}^{\prime}) \cdot \mathbf{X}(\mathbf{r}^{\prime}) g_0(\mathbf{r}, \mathbf{r}^{\prime})\,ds^{\prime}
\end{aligned}
\end{equation}
Here, $\partial V$ is the boundary surface of the volume $V$ and $\hat{\mathbf{n}}(\mathbf{r})$ is the outward unit normal vector on $\partial V$. The scalar Green function of the homogeneous background medium $g_0(\mathbf{r},\mathbf{r}')$ is given by
\begin{equation}
    g_0(\mathbf{r},\mathbf{r}') = \frac{e^{-\mathrm{j} k_0 |\mathbf{r}-\mathbf{r}'|}}{4\pi|\mathbf{r}-\mathbf{r}'|}
    \label{eq:scalar_green}
\end{equation}
where $k_0 = \omega\sqrt{\mu_0\varepsilon_0} = 2\pi/\lambda_0$ is the background wavenumber and $\lambda_0$ is the corresponding wavelength.
The total electric field $\mathbf{E}(\mathbf{r})$, the scattered electric field $\mathbf{E}^{\mathrm{sca}}(\mathbf{r})$, and the incident field $\mathbf{E}^{\mathrm{inc}}(\mathbf{r})$ satisfy the fundamental field relation
\begin{equation}
    \mathbf{E}(\mathbf{r})=\mathbf{E}^{\mathrm{inc}}(\mathbf{r})+\mathbf{E}^{\mathrm{sca}}(\mathbf{r}).
    \label{eq:field_relation}
\end{equation}
In the dielectric region $V_{\mathrm{B}}$, $\mathbf{J}_\mathrm{B}(\mathbf{r})$ represents the bound-electron polarization current density and is related to $\mathbf{E}(\mathbf{r})$ as
\begin{equation}
    \mathbf{J}_\mathrm{B}(\mathbf{r})=\mathrm{j} \omega \varepsilon_0[\varepsilon_\mathrm{B}(\mathbf{r})-1] \mathbf{E}(\mathbf{r}),\, \mathbf{r}\in V_\mathrm{B}
    \label{eq:bound_current}
\end{equation}
where $\varepsilon_\mathrm{B}(\mathbf{r})$ is the relative permittivity of the dielectric region. 

In the metallic region $V_{\mathrm{H}}$, $\mathbf{J}_{\mathrm{H}}(\mathbf{r})$ represents the free-electron current density, referred to here as the hydrodynamic current density. It is related to $\mathbf{E}(\mathbf{r})$ through the HDE 
\begin{equation}
    [\mathrm{j}\omega + \gamma]\mathbf{J}_{\mathrm{H}}(\mathbf{r})+\beta^2 \nabla \rho_\mathrm{H}(\mathbf{r})=\omega_{\mathrm{p}}^2 \varepsilon_0 \mathbf{E}(\mathbf{r}),\, \mathbf{r}\in V_\mathrm{H}.
    \label{eq:hydro_current}
\end{equation}
Here, $\rho_\mathrm{H}(\mathbf{r})$ is the free-electron charge density, $\beta$ is the nonlocal parameter related to the Fermi velocity $v_\mathrm{F}$ by $\beta^2 = 0.6\,v_\mathrm{F}^2$, $\gamma$ is the damping constant, and $\omega_{\mathrm{p}}$ is the plasma frequency of the metal. The continuity equation relates $\rho_\mathrm{H}(\mathbf{r})$ and $\mathbf{J}_{\mathrm{H}}(\mathbf{r})$ as
\begin{equation}
\label{eq:continuity}
\nabla \cdot \mathbf{J}_{\mathrm{H}}(\mathbf{r}) = -\mathrm{j}\omega \rho_\mathrm{H}(\mathbf{r}).
\end{equation}
Using~\eqref{eq:bound_current}, ~\eqref{eq:hydro_current}, and~\eqref{eq:continuity}, $\mathbf{E}(\mathbf{r})$ can be expressed in terms of the unknown current densities $\mathbf{J}_\mathrm{B}(\mathbf{r})$ and $\mathbf{J}_\mathrm{H}(\mathbf{r})$ in $V_\mathrm{B}$ and $V_\mathrm{H}$ as
\begin{equation}
    \mathbf{E}(\mathbf{r}) = \frac{1}{\mathrm{j}\omega \varepsilon_0 [\varepsilon_\mathrm{B}(\mathbf{r})-1]}\mathbf{J}_{\mathrm{B}}(\mathbf{r}),\, \mathbf{r} \in V_\mathrm{B}
\label{eq:efield_bound}
\end{equation}
\begin{equation}
    \mathbf{E}(\mathbf{r}) = \frac{\mathrm{j}\beta^2}{\omega \omega_{\mathrm{p}}^2 \varepsilon_0} \nabla[\nabla \cdot \mathbf{J}_{\mathrm{H}}(\mathbf{r})]  +\frac{\mathrm{j}\omega+ \gamma}{ \omega_{\mathrm{p}}^2 \varepsilon_0} \mathbf{J}_{\mathrm{H}}(\mathbf{r}),\,\mathbf{r} \in V_{\mathrm{H}}.
\label{eq:efield_free}
\end{equation}
Substituting~\eqref{eq:e_sca} for $\mathbf{E}^{\mathrm{sca}}(\mathbf{r})$ and ~\eqref{eq:efield_bound} and~\eqref{eq:efield_free} for $\mathbf{E}(\mathbf{r})$ in~\eqref{eq:field_relation} yields the coupled VIE--HDE system as
\begin{equation}
\label{eq:vie_B}
     \frac{1}{\mathrm{j}\omega \varepsilon_0 [\varepsilon_\mathrm{B}(\mathbf{r})-1]}\mathbf{J}_{\mathrm{B}}(\mathbf{r}) -\mathcal{L}_{V_\mathrm{B}}[\mathbf{J}_\mathrm{B}] (\mathbf{r})-\mathcal{L}_{V_\mathrm{H}}[\mathbf{J}_\mathrm{H}] (\mathbf{r})=\mathbf{E}^{\mathrm{inc}}(\mathbf{r}),\,\mathbf{r} \in V_\mathrm{B}
\end{equation}
\begin{equation}
\label{eq:vie_H}
     -\mathcal{L}_{V_\mathrm{B}}[\mathbf{J}_\mathrm{B}](\mathbf{r})
     + \frac{\mathrm{j}\beta^2}{\omega \omega_{\mathrm{p}}^2 \varepsilon_0}
     \nabla\left[\nabla \cdot \mathbf{J}_{\mathrm{H}}(\mathbf{r})\right]
     +\frac{\mathrm{j}\omega+\gamma}{\omega_{\mathrm{p}}^2 \varepsilon_0}
     \mathbf{J}_{\mathrm{H}}(\mathbf{r})-\mathcal{L}_{V_\mathrm{H}}[\mathbf{J}_\mathrm{H}](\mathbf{r})
     =\mathbf{E}^{\mathrm{inc}}(\mathbf{r}),\quad
     \mathbf{r}\in V_\mathrm{H}.
\end{equation}
This coupled system of equations is complemented by the hard-wall boundary condition at the interfaces between metallic and dielectric regions~\cite{boardman1981boundary}
\begin{equation}
    \hat{\mathbf{n}}_{\mathrm{H}}(\mathbf{r}) \cdot \mathbf{J}_{\mathrm{H}}(\mathbf{r}) = 0,\, \mathbf{r} \in \partial V_\mathrm{H}
    \label{eq:bc}
\end{equation}
 where $\partial V_\mathrm{H}$ is the boundary surface of $V_{\mathrm{H}}$ and $\hat{\mathbf{n}}_{\mathrm{H}}(\mathbf{r})$  is the outward unit normal vector on $\partial V_\mathrm{H}$. This boundary condition states that there is no free-electron flow across the metal boundary, and it is enforced through the selection of the basis functions used to expand $\mathbf{J}_{\mathrm{H}}(\mathbf{r})$, as described in the next section. 
 
\subsection{Discretization}\label{sec:discretization}
To discretize the coupled VIE--HDE system in~\eqref{eq:vie_B} and~\eqref{eq:vie_H}, the volume $V_\mathrm{C}$ is partitioned into a mesh of tetrahedral elements, and the unknown current densities $\mathbf{J}_\mathrm{B}(\mathbf{r})$ and $\mathbf{J}_\mathrm{H}(\mathbf{r})$ are expanded using Schaubert-Wilton-Glisson (SWG) basis functions~\cite{schaubert1984tetrahedral} as
\begin{equation}
\label{eq:current_expansion}
\begin{aligned}
    \mathbf{J}_{\mathrm{B}} (\mathbf{r}) = \sum_{n=1}^{N_\mathrm{B}} \{ \bar{I}_\mathrm{B}\}_n \mathbf{f}^\mathrm{B}_n(\mathbf{r}),\,\mathbf{r} \in {V}_\mathrm{B}\\
    \mathbf{J}_{\mathrm{H}} (\mathbf{r}) = \sum_{n=1}^{N_\mathrm{H}} \{ \bar{I}_\mathrm{H}\}_n \mathbf{f}^\mathrm{H}_n(\mathbf{r}),\,\mathbf{r} \in {V}_\mathrm{H}.
    \end{aligned}
\end{equation}
Here, $\bar{I}_\mathrm{B}$ and $\bar{I}_\mathrm{H}$ are the vectors storing the unknown expansion coefficients, and $N_\mathrm{B}$ and $N_\mathrm{H}$ denote the numbers of basis functions used to expand $\mathbf{J}_\mathrm{B}(\mathbf{r})$ and $\mathbf{J}_\mathrm{H}(\mathbf{r})$, respectively. The SWG basis function associated with the $n$th triangular face $S_n$ is defined as
\begin{equation}
    \mathbf{f}_n(\mathbf{r})=\begin{cases}
\displaystyle\mathbf{f}_n^{+}(\mathbf{r})=\frac{|S_n|}{3|V^+_\mathrm{n}|}(\mathbf{r}-\mathbf{r}_n^{+}),\,\mathbf{r} \in {V^{+}_n} \\
\displaystyle\mathbf{f}_n^{-}(\mathbf{r})=-\frac{|S_n|}{3|V^-_\mathrm{n}|}(\mathbf{r}-\mathbf{r}_n^{-}),\,\mathbf{r} \in {V^{-}_n} \\
0,\,\mathrm{elsewhere}
\end{cases}.
\label{eq:swg}
\end{equation}
Here, $V_n^{+}$ and $V_n^{-}$ are the two tetrahedra sharing $S_n$, $|S_n|$ is the area of $S_n$, $|V_n^{+}|$ and $|V_n^{-}|$ are the volumes of $V_n^{+}$ and $V_n^{-}$, and $\mathbf{r}_n^{+}$ and $\mathbf{r}_n^{-}$ are the positions of the free nodes of $V_n^{+}$ and $V_n^{-}$ opposite $S_n$, respectively. 

The bound-electron current density $\mathbf{J}_\mathrm{B}(\mathbf{r})$ is expanded using a combination of full- and half-SWG basis functions. The full-SWG functions are associated with the internal faces of the mesh partitioning $V_\mathrm{B}$, whereas the half-SWG functions are associated with the triangular faces on the boundary surface $\partial V_\mathrm{B}$. Since a boundary face is adjacent to only one tetrahedron, its half-SWG function consists of the single branch $\mathbf{f}_n^{+}(\mathbf{r})$ supported on that tetrahedron. The normal component of such a function does not vanish on $\partial V_\mathrm{B}$~\cite{zhang2015solving}. In contrast, the free-electron current density $\mathbf{J}_\mathrm{H}(\mathbf{r})$ is expanded using only full-SWG basis functions. The hard-wall boundary condition in~\eqref{eq:bc} requires the normal component of $\mathbf{J}_\mathrm{H}(\mathbf{r})$ to vanish on $\partial V_\mathrm{H}$, which is enforced automatically by full-SWG functions, since their normal component vanishes on the boundary~\cite{uulu2023coupled}.

Substituting the expansions in~\eqref{eq:current_expansion} into the coupled VIE--HDE system in~\eqref{eq:vie_B} and~\eqref{eq:vie_H}, and applying the Galerkin scheme with the testing functions $\mathbf{f}_{m}^\mathrm{B}(\mathbf{r})$, $m=1,2,\ldots,N_\mathrm{B}$, and $\mathbf{f}_m^\mathrm{H}(\mathbf{r})$, $m=1,2,\ldots,N_\mathrm{H}$,
yields the coupled matrix system of dimension $(N_\mathrm{B}+N_\mathrm{H}) \times (N_\mathrm{B}+N_\mathrm{H})$, given by
\begin{equation}
\underbrace{\begin{bmatrix}
\bar{\bar{Z}}_{\mathrm{BB}} & \bar{\bar{Z}}_{\mathrm{BH}} \\
\bar{\bar{Z}}_{\mathrm{HB}} & \bar{\bar{Z}}_{\mathrm{HH}}
\end{bmatrix}}_{\displaystyle \bar{\bar{Z}}} \underbrace{\begin{bmatrix}
\bar{I}_{\mathrm{B}} \\
\bar{I}_{\mathrm{H}}
\end{bmatrix}}_{\displaystyle \bar{I}}=\underbrace{\begin{bmatrix}
\bar{V}^{\mathrm{inc}}_\mathrm{B} \\
\bar{V}^{\mathrm{inc}}_\mathrm{H}
\end{bmatrix}}_{\displaystyle \bar{V}^\mathrm{inc}}.
\label{eq:coupled_matrix}
\end{equation}
The entries of each matrix block are
\begin{equation}
\{\bar{\bar{Z}}_{\mathrm{BB}}\}_{mn}=\frac{1}{\mathrm{j}\omega \varepsilon_0}\big\langle\mathbf{f}_m^{\mathrm{B}}(\mathbf{r}), \frac{\mathbf{f}_n^{\mathrm{B}}(\mathbf{r})}{\varepsilon_{\mathrm{B}}(\mathbf{r})-1}\big\rangle-\big\langle\mathbf{f}_m^{\mathrm{B}}(\mathbf{r}), \mathcal{L}_{V_{\mathrm{B}}}[\mathbf{f}_n^{\mathrm{B}}](\mathbf{r})\big\rangle,\, m,n=1,2,\ldots,N_\mathrm{B}
\end{equation}
\begin{equation}
\{\bar{\bar{Z}}_{\mathrm{BH}}\}_{mn}=-\big\langle\mathbf{f}_m^{\mathrm{B}}(\mathbf{r}), \mathcal{L}_{V_{\mathrm{H}}}[\mathbf{f}_n^{\mathrm{H}}](\mathbf{r})\big\rangle, \, m=1,2,\ldots,N_\mathrm{B},\, n=1,2,\ldots,N_\mathrm{H}
\end{equation}
\begin{equation}
\{\bar{\bar{Z}}_{\mathrm{HB}}\}_{mn}=-\big\langle\mathbf{f}_m^{\mathrm{H}}(\mathbf{r}), \mathcal{L}_{V_{\mathrm{B}}}[\mathbf{f}_n^{\mathrm{B}}](\mathbf{r})\big\rangle,\,m=1,2,\ldots,N_\mathrm{H},\,n=1,2,\ldots,N_\mathrm{B}
\end{equation}
\begin{equation}
\begin{aligned}
\{\bar{\bar{Z}}_{\mathrm{HH}}\}_{mn}=\frac{\mathrm{j}\beta^2}{\omega\omega_{\mathrm{p}}^2\varepsilon_0}\big\langle\mathbf{f}_m^{\mathrm{H}}(\mathbf{r}), \nabla[\nabla \cdot \mathbf{f}_n^{\mathrm{H}}(\mathbf{r})]\big\rangle
+\frac{\mathrm{j}\omega+ \gamma}{ \omega_{\mathrm{p}}^2\varepsilon_0}\big\langle\mathbf{f}_m^{\mathrm{H}}(\mathbf{r}), \mathbf{f}_n^{\mathrm{H}}(\mathbf{r})\big\rangle-\big\langle\mathbf{f}_m^{\mathrm{H}}(\mathbf{r}), \mathcal{L}_{V_{\mathrm{H}}}[\mathbf{f}_n^{\mathrm{H}}](\mathbf{r})\big\rangle,\\
m,n=1,2,\ldots,N_\mathrm{H}
\end{aligned}
\end{equation}
and the right-hand side vectors are
\begin{equation}
\{\bar{V}^{\mathrm{inc}}_\mathrm{B}\}_{m} = \big\langle \mathbf{f}_m^{\mathrm{B}}(\mathbf{r}), \mathbf{E}^{\mathrm{inc}}(\mathbf{r}) \big\rangle,\ m=1,2,\ldots,N_\mathrm{B}
\end{equation}
\begin{equation}
\{\bar{V}^{\mathrm{inc}}_\mathrm{H}\}_{m} = \big\langle\mathbf{f}_m^{\mathrm{H}}(\mathbf{r}), \mathbf{E}^{\mathrm{inc}}(\mathbf{r}) \big\rangle,\ m=1,2,\ldots,N_\mathrm{H}.
\end{equation}
Here, the inner product between two vector functions $\mathbf{a}(\mathbf{r})$ and $\mathbf{b}(\mathbf{r})$ is defined over $V_a$, the volumetric support of $\mathbf{a}(\mathbf{r})$, as
\begin{equation}
    \big\langle\mathbf{a}(\mathbf{r}), \mathbf{b}(\mathbf{r})\big\rangle = \int_{V_a} \mathbf{a}(\mathbf{r}) \cdot \mathbf{b}(\mathbf{r})\,dv.
    \label{eq:innerprod}
\end{equation}
The explicit expressions of the matrix entries and the right-hand side vectors are provided in Appendix~\ref{appendix:C}. 

The discretized coupled VIE--HDE system in~\eqref{eq:coupled_matrix} is used in the CMA formulations discussed in the next section.
                    
\subsection{Full-Structure CMA Formulation}
\label{sec:full_cma_section}
CMA obtains the characteristic modes as the eigensolutions of the GEE~\cite{harrington1971theory, harrington1971computation, ylaoijala2019generalized}
\begin{equation}
    \bar{\bar{Z}}\bar{I}_k=(1+\mathrm{j}\lambda_k) \bar{\bar{W}} \bar{I}_k
    \label{eq:gee}
\end{equation}
where $\bar{I}_k=[\bar{I}^{\top}_{\mathrm{B}_k}\,\bar{I}^{\top}_{\mathrm{H}_k}]^\top$ is the coefficient vector of the $k$th characteristic current, and $\lambda_k$ is the corresponding characteristic eigenvalue. Its subvectors $\bar{I}_{\mathrm{B}_k}$ and $\bar{I}_{\mathrm{H}_k}$ contain the expansion coefficients of the bound-electron and free-electron characteristic currents, respectively. Here, $\bar{\bar{Z}}$ is the matrix in~\eqref{eq:coupled_matrix} arising from the discretization of the coupled VIE--HDE system in ~\eqref{eq:vie_B} and~\eqref{eq:vie_H}, and $\bar{\bar{W}}$ is a weighting matrix that determines which resonance (extinction, radiation, or absorption) the GEE characterizes. In this work, extinction resonances are considered, and the weighting matrix is chosen as $\bar{\bar{W}}=\bar{\bar{R}}$, where $\bar{\bar{R}} = \mathrm{Re}\{\bar{\bar{Z}}\}$ is associated with extinction power. The physical interpretation of this choice is detailed in Appendix~\ref{appendix:A}.

Accordingly, the GEE takes the form
\begin{equation}
    \bar{\bar{Z}}\bar{I}_k=(\bar{\bar{R}}+\mathrm{j}\bar{\bar{X}})\bar{I}_k=(1+\mathrm{j}\lambda_k)\,\bar{\bar{R}}\,\bar{I}_k
    \label{eq:gee_split}
\end{equation}
where $\bar{\bar{X}} = \mathrm{Im}\{\bar{\bar{Z}}\}$, which gives
\begin{equation}
    \bar{\bar{X}}\bar{I}_k=\lambda_k \bar{\bar{R}} \bar{I}_k.
    \label{eq:gee2}
\end{equation}
This formulation is referred to as the full-structure CMA, since the resulting characteristic modes are supported on the entire composite structure $V_\mathrm{C}=V_\mathrm{B}\cup V_\mathrm{H}$. Consequently, characteristic currents with eigenvalues close to $0$ represent the resonant modes of the composite nanostructure, whether localized in the metallic region, in the dielectric region, or distributed across both. Since the GEE is real-symmetric, the characteristic currents are real, or equiphase. As a normalized and more intuitive alternative to the eigenvalue, the modal significance (MS) is defined as~\cite{harrington1971computation, chen2015characteristic}
\begin{equation}
     \sigma_k= \frac{1}{|1+\mathrm{j}\lambda_k|}.
     \label{eq:ms}
\end{equation}
A mode is at resonance when its MS approaches $1$. MS is therefore adopted as the primary indicator of resonance in the numerical examples.

\subsection{Sub-Structure CMA Formulation}
In this section, the sub-structure CMA formulation~\cite{alroughani2014classification,alroughani2016orthogonality} is derived from the reduced form of the discretized coupled VIE--HDE system~\eqref{eq:coupled_matrix}. By eliminating the dielectric-region current unknowns, this formulation constructs the GEE using only the metallic-region current unknowns, while the effect of the dielectric region is retained through the reduced matrix. This allows the resonant modes associated with the metallic region to be isolated. Eliminating $\bar{I}_\mathrm{B}$ from the coupled system~\eqref{eq:coupled_matrix} through the Schur complement yields the reduced system
\begin{equation}
    \bar{\bar{Z}}_\mathrm{sub} \bar{I}_\mathrm{H} = \bar{V}_\mathrm{sub}^\mathrm{inc}
    \label{eq:reduced_matrix}
\end{equation}
where the sub-structure matrix $\bar{\bar{Z}}_\mathrm{sub}$ is given by 
\begin{equation}
    \bar{\bar{Z}}_\mathrm{sub} = \bar{\bar{Z}}_\mathrm{HH}-\bar{\bar{Z}}_\mathrm{HB}\bar{\bar{Z}}_\mathrm{BB}^{-1}\bar{\bar{Z}}_\mathrm{BH} 
    \label{eq:z_sub}
\end{equation}
and the reduced excitation vector $\bar{V}_\mathrm{sub}^\mathrm{inc}$ is given by
\begin{equation}
    \bar{V}_\mathrm{sub}^\mathrm{inc} = \bar{V}_\mathrm{H}^\mathrm{inc}-\bar{\bar{Z}}_\mathrm{HB}\bar{\bar{Z}}_\mathrm{BB}^{-1}\bar{V}_\mathrm{B}^\mathrm{inc}.
    \label{eq:v_sub}
\end{equation}
Following the same construction as the full-structure CMA, the GEE for the sub-structure CMA is
\begin{equation}
    \bar{\bar{X}}_\mathrm{sub} \bar{I}_{\mathrm{H}_k}=\lambda_{\mathrm{sub}_k} \bar{\bar{R}}_\mathrm{sub} \bar{I}_{\mathrm{H}_k}
\label{eq:gee_sub}
\end{equation}
where $\bar{\bar{R}}_\mathrm{sub}=\mathrm{Re}\{\bar{\bar{Z}}_\mathrm{sub}\}$ and $\bar{\bar{X}}_\mathrm{sub}=\mathrm{Im}\{\bar{\bar{Z}}_\mathrm{sub}\}$, $\bar{I}_{\mathrm{H}_k}$ is the coefficient vector of the $k$th sub-structure characteristic current, and $\lambda_{\mathrm{sub}_k}$ is the corresponding sub-structure characteristic eigenvalue. The physical interpretation of the sub-structure formulation is detailed in Appendix~\ref{appendix:B}. Modes with eigenvalues $\lambda_{\mathrm{sub}_k}$ close to $0$ are associated with resonances. As in the full-structure CMA, the real-symmetric form of the GEE yields real, or equiphase, characteristic currents; here, however, they are confined to the metallic region~\cite{xiang2019preliminary}.

One possible way to recover the characteristic currents associated with the dielectric region from the metallic-region characteristic currents is through the original coupled VIE--HDE system~\eqref{eq:coupled_matrix}. However, this yields complex-valued currents, making direct comparison with the real currents of the full-structure CMA impractical. To address this issue, the dielectric-region characteristic currents are instead recovered from the first block row of the full-structure GEE~\eqref{eq:gee2}, expressed in terms of the metallic-region currents as
\begin{equation}
    \bar{I}_{\mathrm{B}_k} =(\bar{\bar{X}}_{\mathrm{BB}} - \lambda_{\mathrm{sub}_k} \bar{\bar{R}}_{\mathrm{BB}})^{-1} \left(-\bar{\bar{X}}_{\mathrm{BH}} + \lambda_{\mathrm{sub}_k} \bar{\bar{R}}_{\mathrm{BH}}\right)    \bar{I}_{\mathrm{H}_k}
    \label{eq:curD_mode_sub}
\end{equation}
where $\bar{\bar{R}}_{\mathrm{BB}}=\mathrm{Re}\{\bar{\bar{Z}}_{\mathrm{BB}}\}$, $\bar{\bar{X}}_{\mathrm{BB}}=\mathrm{Im}\{\bar{\bar{Z}}_{\mathrm{BB}}\}$, $\bar{\bar{R}}_{\mathrm{BH}}=\mathrm{Re}\{\bar{\bar{Z}}_{\mathrm{BH}}\}$, and $\bar{\bar{X}}_{\mathrm{BH}}=\mathrm{Im}\{\bar{\bar{Z}}_{\mathrm{BH}}\}$. This ensures that $\bar{I}_{\mathrm{B}_k}$ are real or equiphase, enabling a consistent and meaningful comparison between the full-structure and sub-structure CMA results. 

\subsection{Computational Complexity}
Each GEE is solved using the implicitly restarted Arnoldi method (IRAM)~\cite{sorensen1992implicit} implemented in the ARPACK library~\cite{lehoucq1998arpack}. In the full-structure CMA, the GEE in~\eqref{eq:gee} has dimension $N=N_\mathrm{B}+N_\mathrm{H}$ and is first transformed into the standard eigenvalue equation
\begin{equation}
    \bar{\bar{X}}^{-1}\bar{\bar{R}} \bar{I}_k=v_k\bar{I}_k
    \label{eq:standard_ee}
\end{equation}
where $v_k=1/\lambda_k$. The equation is then solved for a small number $n_\mathrm{ev}$ of the largest eigenvalues, which correspond to the smallest eigenvalues of~\eqref{eq:gee}. At each IRAM iteration, the operation $\bar{w} =\bar{\bar{X}}^{-1}\bar{\bar{R}}\bar{z}$ is required. This operation is performed in two steps: computing $\bar{y}=\bar{\bar{R}}\bar{z}$ which costs $\mathcal{O}(N^2)$, and solving $\bar{\bar{X}}\bar{w}=\bar{y}$ iteratively using the transpose-free quasi-minimal residual (TFQMR) method~\cite{freund1993}, which costs $n_\mathrm{it}\,\mathcal{O}(N^2)$. Here, $n_\mathrm{it}$ denotes the number of iterations required for convergence. Thus, if the number of IRAM restart cycles required for convergence is $n_\mathrm{cyc}$, the total cost of solving the full-structure GEE scales as $mn_\mathrm{cyc}(1+n_\mathrm{it})\,\mathcal{O}(N^2)$, where $m$ is the number of Arnoldi basis vectors used in IRAM, with $m\approx 2n_\mathrm{ev}$. 

Similarly, solving the reduced GEE in~\eqref{eq:gee_sub} for the sub-structure CMA costs $mn_\mathrm{cyc}(1+n_\mathrm{it})\mathcal{O}(N_\mathrm{H}^2)$. Hence, the eigensolution stage of the sub-structure CMA is significantly cheaper than that of the full-structure CMA. However, this reduction comes with an additional cost: the reduced matrix~\eqref{eq:z_sub} must be constructed before solving the reduced GEE. Its construction requires the $\mathcal{O}(N_\mathrm{B}^3)$ direct factorization of $\bar{\bar{Z}}_\mathrm{BB}$, followed by solving the multiple-right-hand-side system $\bar{\bar{Z}}_\mathrm{BB}\bar{\bar{Y}}=\bar{\bar{Z}}_\mathrm{BH}$, whose $N_\mathrm{H}$ right-hand sides are handled by back-substitution against the existing factorization at a cost of $\mathcal{O}(N_\mathrm{B}^2 N_\mathrm{H})$, and the multiplication of $\bar{\bar{Z}}_\mathrm{HB}$ by $\bar{\bar{Y}}$ at a cost of $\mathcal{O}(N_\mathrm{B}N_\mathrm{H}^2)$, giving a total construction cost of $\mathcal{O}(N_\mathrm{B}^3+N_\mathrm{B}^2 N_\mathrm{H}+N_\mathrm{B}N_\mathrm{H}^2)$.  Therefore, the sub-structure CMA is computationally advantageous only when the reduction in the eigenvalue problem size outweighs the additional cost of constructing the reduced matrix.

\section{Numerical Results}
\label{sec:num_results}
In this section, several numerical examples are presented to investigate the resonance behavior of metallic nanostructures in the presence of dielectric media. The results are validated against the extinction cross section (ECS), which is computed under plane-wave excitation. The background medium is vacuum, and the incident electric field of this plane wave is given by
\begin{equation}
    \mathbf{E}^{\mathrm{inc}}(\mathbf{r})=\hat{\mathbf{p}}E_0 e^{-\mathrm{j}k_0\hat{\mathbf{k}}\cdot\mathbf{r}}
    \label{eq:plane_wave}
\end{equation}
where $E_0=1\,\mathrm{V/m}$ is the amplitude, $\hat{\mathbf{p}}$ is the polarization unit vector, and $\hat{\mathbf{k}}$ is the unit vector in the direction of propagation. 

For the isolated metallic nanostructures, the ECS is computed using the HDVIE solver of~\cite{khamitova2024characteristic, khamitova2026characteristic}, and their modal analysis follows the CMA formulation presented therein. For the composite nanostructures, the ECS is computed by solving the discretized coupled VIE--HDE system in~\eqref{eq:coupled_matrix}, and the full-structure and sub-structure CMA formulations are used, individually or together. In all examples, the ten dominant characteristic currents are computed, and the resonances identified by the MS curves are validated against the corresponding ECS resonances.

As demonstrated in Section~\ref{sec:nanorod}, when the dielectric does not support resonances in the frequency range of interest, all resonances in this range are of metallic origin, and the full- and sub-structure formulations identify the same set of resonance frequencies. In this regime, the sub-structure formulation is preferred, as it is computationally more efficient.

\subsection{Composite Nanosphere}\label{sec:nanosphere}
\begin{figure}[ht!]
\centering
  \includegraphics[width=0.4\columnwidth]{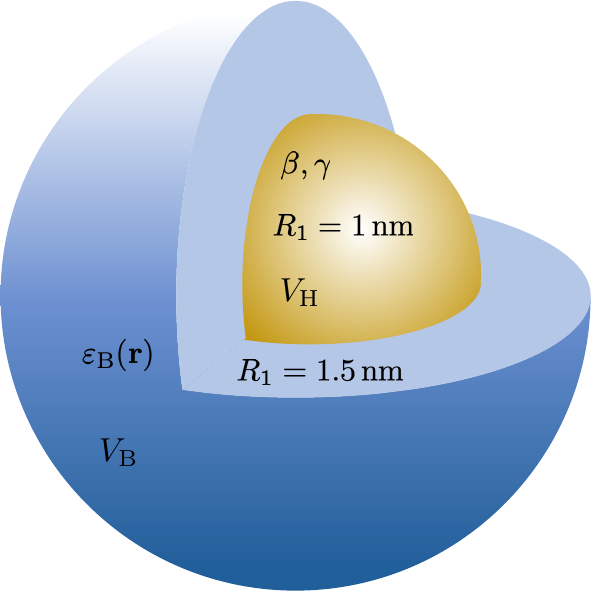}
  \caption{A silica-coated gold nanosphere.}\label{fig:ex1_problem}
\end{figure}

This example investigates the resonance behavior of a gold nanosphere and the effect of a silica coating on its resonance characteristics, as shown in Fig.~\ref{fig:ex1_problem}. The gold nanosphere has a radius of $1\, \mathrm{nm}$, and the spherical silica shell has a thickness of $0.5\, \mathrm{nm}$. The hydrodynamic parameters for gold are $\omega_{\mathrm{p}}=1.20\times10^{16}\, \mathrm{rad/s}$, $\gamma=1.36\times10^{14}\, \mathrm{rad/s}$, and $v_\mathrm{F}=1.39\times10^{6}\, \mathrm{m/s}$~\cite{uulu2023coupled}, and the relative permittivity of silica is $\varepsilon_\mathrm{B}(\mathbf{r}) = 2.25$. Simulations are carried out in the frequency range $[0.5,1.2]\,\omega_{\mathrm{p}}$. For the plane-wave excitation, the polarization and propagation directions are $\hat{\mathbf{p}}=\hat{\mathbf{x}}$ and $\hat{\mathbf{k}}=\hat{\mathbf{z}}$. Two cases are considered: the isolated gold nanosphere and the composite nanosphere. For the composite nanosphere, the full-structure CMA formulation is used. For both cases, two meshes are used: a coarser mesh over the frequency range $[0.5,1.0]\,\omega_{\mathrm{p}}$ and a denser mesh over $[1.0,1.2]\,\omega_{\mathrm{p}}$ to improve accuracy at higher frequencies. For the low-frequency simulations, the gold nanosphere is discretized using a mesh with $8\, 619$ unknowns, while the composite nanosphere is discretized using a mesh with $N_\mathrm{B} = 12\, 974$ and $N_\mathrm{H} = 8\, 756$ unknowns. For the high-frequency simulations, the gold nanosphere is discretized using a mesh with $105\, 233$ unknowns, while the composite nanosphere is discretized using a mesh with $N_\mathrm{B} = 47\, 246$ and $N_\mathrm{H} = 105\, 374$ unknowns. 

\begin{figure}[ht!]
    \centering
        \subfloat[]{\includegraphics[width=0.6\columnwidth]{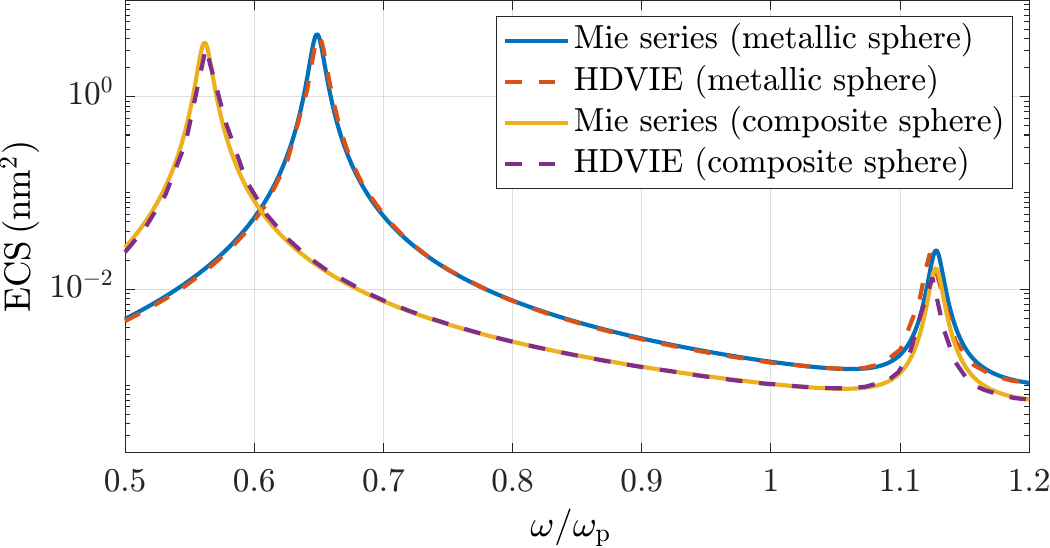}}\\
        \subfloat[]{\includegraphics[width=0.6\columnwidth]{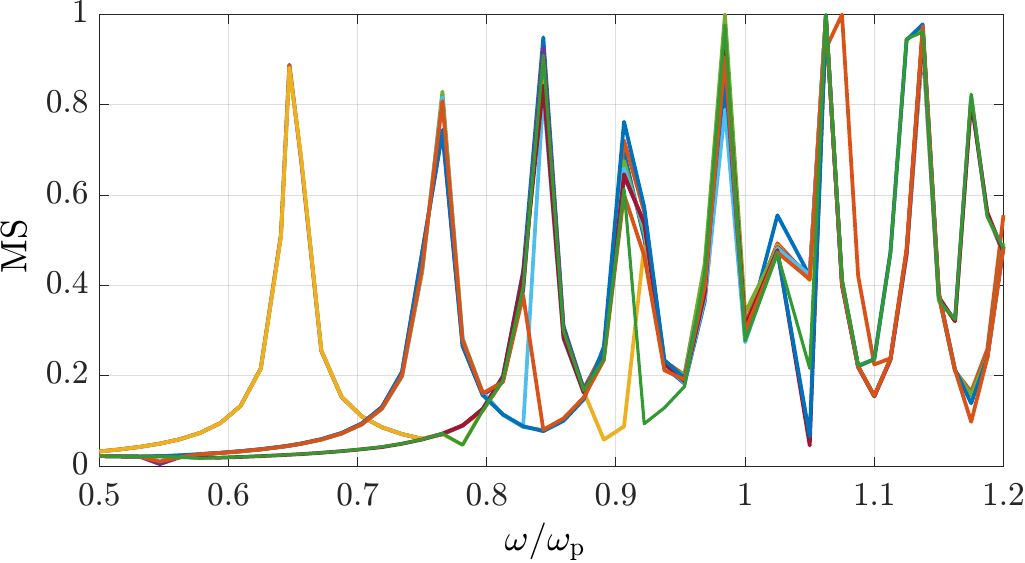}}\\
        \subfloat[]{\includegraphics[width=0.6\columnwidth]{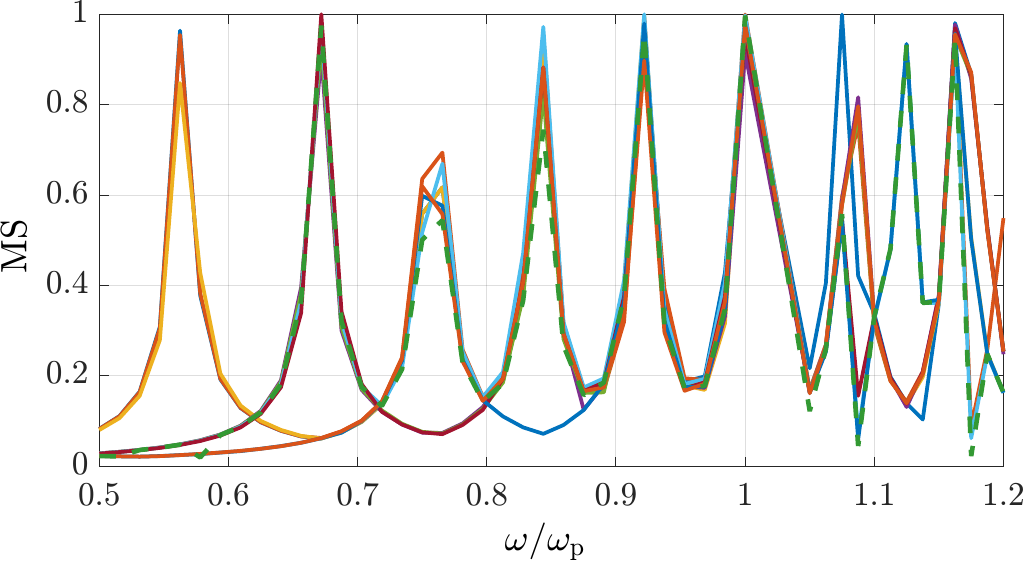}}
    \caption{(a) ECS of the isolated gold and composite nanospheres computed numerically and validated against the nonlocal Mie series. MS curves of (b) the isolated gold nanosphere and (c) the composite nanosphere obtained using the full-structure CMA.}
    \label{fig:ecs_ms_ex1}
\end{figure}

Fig.~\ref{fig:ecs_ms_ex1}(a) compares the resulting ECS of the isolated gold nanosphere and the composite nanosphere with the corresponding nonlocal Mie series~\cite{christensen2014nonlocal,zouros2020monitoring}. The two sets of results are in very good agreement, validating the accuracy of the computed ECS. It can be observed that the addition of the silica coating red-shifts the transverse resonance peak, whereas the longitudinal resonance peak, which arises from the nonlocal pressure wave and appears above the plasma frequency, remains nearly unchanged. Fig.~\ref{fig:ecs_ms_ex1}(b) and~\ref{fig:ecs_ms_ex1}(c) show the MS curves of the isolated gold nanosphere and the composite nanosphere, respectively. The resonances observed in the MS curves agree well with the corresponding ECS resonances, confirming the validity of the CMA formulation. Some additional resonances appear in the MS curves but are not observed in the ECS response, which can be explained by the weak coupling between these resonant modes and the excitation source~\cite{khamitova2024characteristic, khamitova2026characteristic}.

\begin{figure}[ht!]
\centering
        \subfloat[]{\includegraphics[width=0.6\columnwidth]{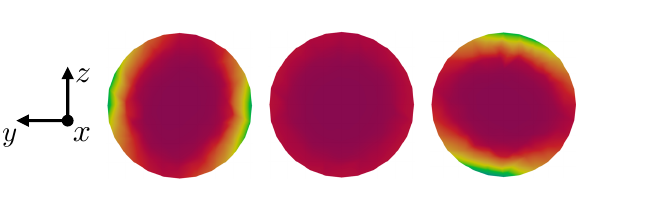}}\\
        \subfloat[]{\includegraphics[width=0.6\columnwidth]{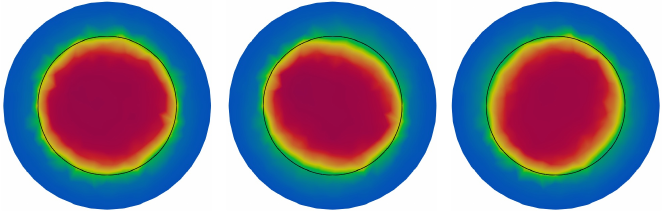}}
    \caption{Visualization of the dominant characteristic currents for (a) the isolated gold nanosphere at $\omega=0.65\, \omega_{\mathrm{p}}$ and (b) the composite nanosphere at $\omega=0.56\, \omega_{\mathrm{p}}$.}
    \label{fig:vis_transverse_ex1}
\end{figure}

\begin{figure}[ht!]
    \centering
        \subfloat[]{\includegraphics[width=0.6\columnwidth]{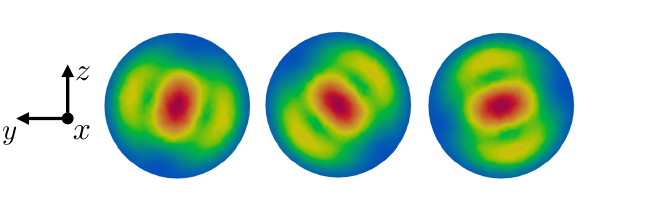}}\\
        \subfloat[]{\includegraphics[width=0.6\columnwidth]{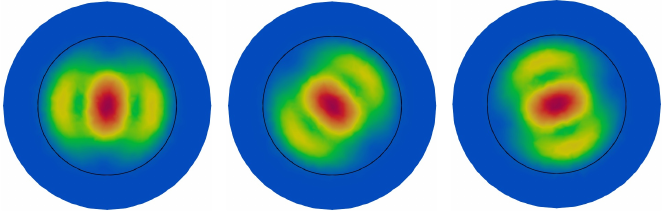}}
    \caption{Visualization of the dominant characteristic currents for (a) the isolated gold nanosphere and (b) the composite nanosphere at $\omega=1.13\, \omega_{\mathrm{p}}$.}
    \label{fig:vis_longitudinal_ex1}
\end{figure}

Fig.~\ref{fig:vis_transverse_ex1} shows the dominant characteristic currents at the first resonance peak, corresponding to the transverse resonance, for the isolated gold nanosphere at $\omega=0.65\, \omega_{\mathrm{p}}$ and the composite nanosphere at $\omega=0.56\, \omega_{\mathrm{p}}$. The results indicate that the silica coating does not significantly alter the resonant current distribution of the gold nanosphere, but shifts the resonance to a lower frequency. Fig.~\ref{fig:vis_longitudinal_ex1} shows the corresponding current distributions at the longitudinal resonance, $\omega=1.13\,\omega_{\mathrm{p}}$. As expected from the ECS results, the resonant current distribution of the gold nanosphere remains essentially unchanged after the addition of the silica coating. 

\subsection{Metallic Nanorod with a Dielectric Slab}\label{sec:nanorod}
\begin{figure}[ht!]
\centering
  \includegraphics[width=0.5\columnwidth]{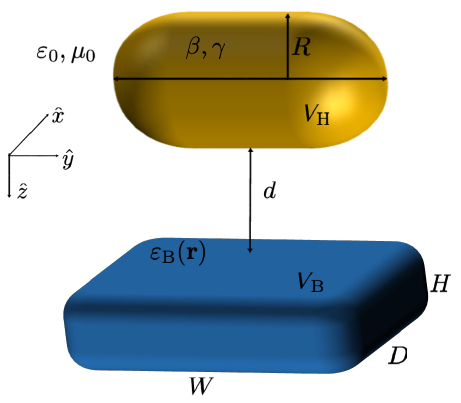}
  \caption{Composite structure comprising a gold nanorod and a silica slab.}
  \label{fig:ex2_problem}
\end{figure}

In this example, the resonance behavior of a gold nanorod is analyzed in the presence of a silica slab, as shown in Fig.~\ref{fig:ex2_problem}. The metallic nanorod has a radius of $1\,\mathrm{nm}$ and a length of $3\,\mathrm{nm}$. The dielectric slab has dimensions of $D = 3\,\mathrm{nm}$, $H = 1\,\mathrm{nm}$, $W= 3\,\mathrm{nm}$, and is separated from the metallic nanorod by a minimum distance of $d =1\,\mathrm{nm}$. As in fabricated structures at nanoscale dimensions, all sharp edges of the slab are smoothed using cylindrical and spherical rounding elements. This also helps to avoid geometrical singularities. The hydrodynamic parameters of the metallic nanorod are the same as those used in Section~\ref{sec:nanosphere}, and the relative permittivity of the slab is $\varepsilon_\mathrm{B}(\mathbf{r})=2.25$. Simulations are carried out in the frequency range $[0.5,1.0]\,\omega_{\mathrm{p}}$. For the plane-wave excitation, the polarization and propagation directions are $\hat{\mathbf{p}}=\hat{\mathbf{x}}$ and $\hat{\mathbf{k}}=\hat{\mathbf{z}}$. Two cases are considered: the isolated metallic nanorod and the composite structure comprising the nanorod and the silica slab. For the composite structure, the full-structure and sub-structure CMA formulations are both used. The isolated metallic nanorod is discretized using a mesh with $29\,454$ unknowns, while the composite structure is discretized using a mesh with $N_\mathrm{B}=11\,652$ and $N_\mathrm{H}=29\,454$ unknowns.  

\begin{figure}[ht!]
        \centering       
        \subfloat[]{\includegraphics[width=0.6\columnwidth]{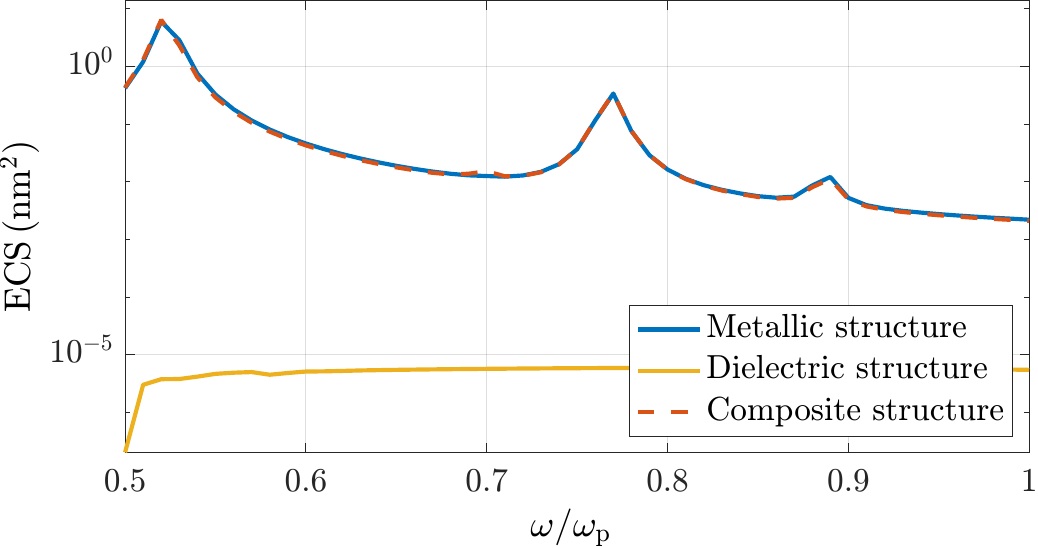}}\\
        \subfloat[]{\includegraphics[width=0.6\columnwidth]{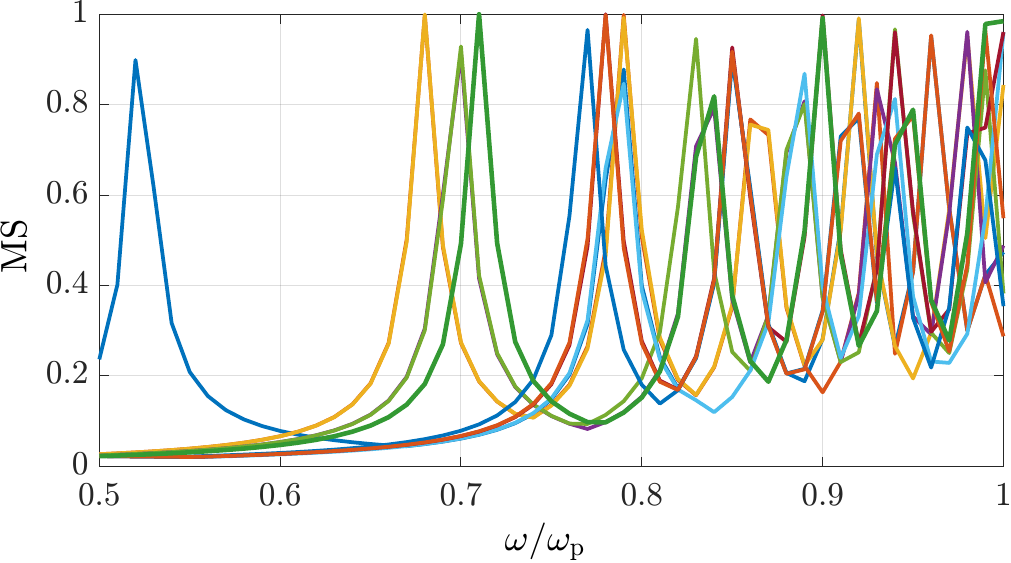}}\\
        \subfloat[]{\includegraphics[width=0.6\columnwidth]{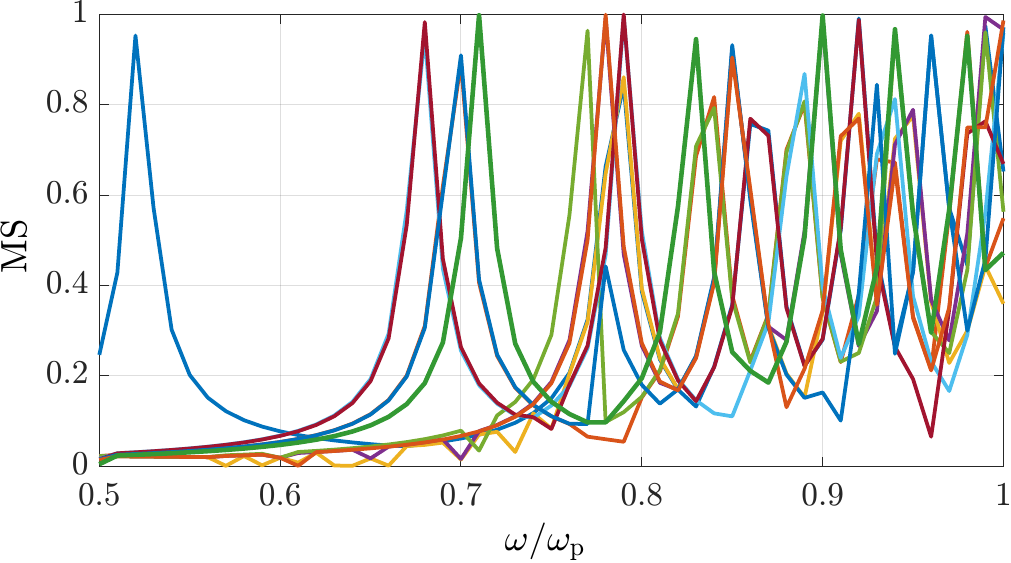}}
    \caption{(a) ECS of the isolated metallic nanorod, the isolated dielectric slab, and the composite nanostructure. MS curves of (b) the isolated metallic nanorod and (c) the composite nanostructure obtained using the full-structure CMA.}
    \label{fig:ecs_ms_ex2}
\end{figure}

Fig.~\ref{fig:ecs_ms_ex2}(a) shows the ECS of the isolated metallic nanorod, the isolated dielectric slab, and the composite nanostructure. The ECS of the dielectric slab is significantly lower than that of the metallic nanorod over the considered frequency range. In addition, the ECS curves of the isolated nanorod and the composite structure nearly overlap, indicating that the resonances of the composite nanostructure are predominantly associated with the metallic nanorod. To quantify the influence of the dielectric slab on the metallic resonances, the MS curves of the isolated metallic nanorod and the composite nanostructure are shown in Fig.~\ref{fig:ecs_ms_ex2}(b) and~\ref{fig:ecs_ms_ex2}(c), respectively, where the composite nanostructure is analyzed using the full-structure CMA. The corresponding resonance frequencies are in close agreement, demonstrating that the dielectric slab only weakly perturbs the resonant behavior of the metallic nanorod.

\begin{figure}[ht!]
    \centering
        \subfloat[]{\includegraphics[width=0.6\columnwidth]{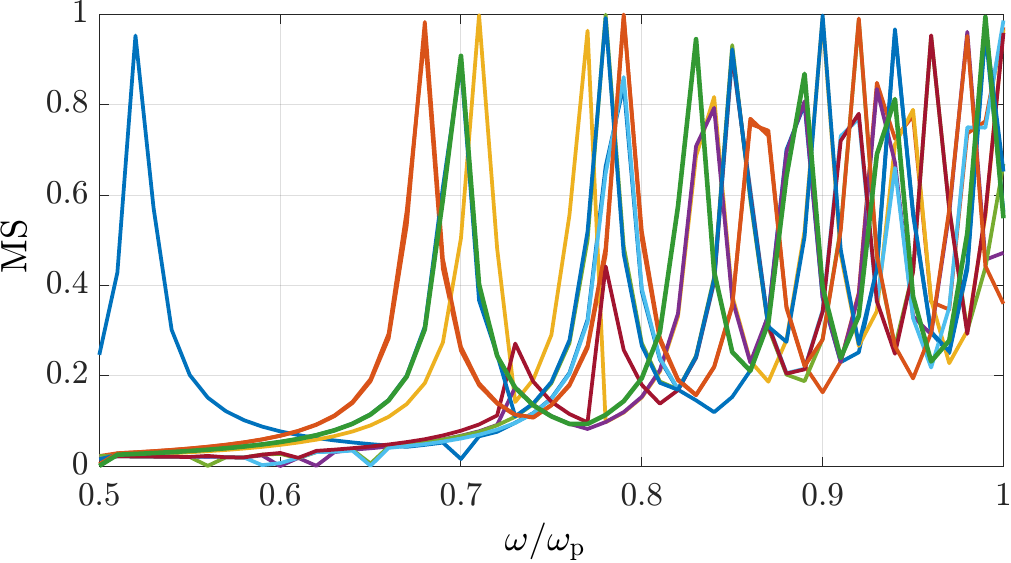}}\\
        \subfloat[]{\includegraphics[width=0.6\columnwidth]{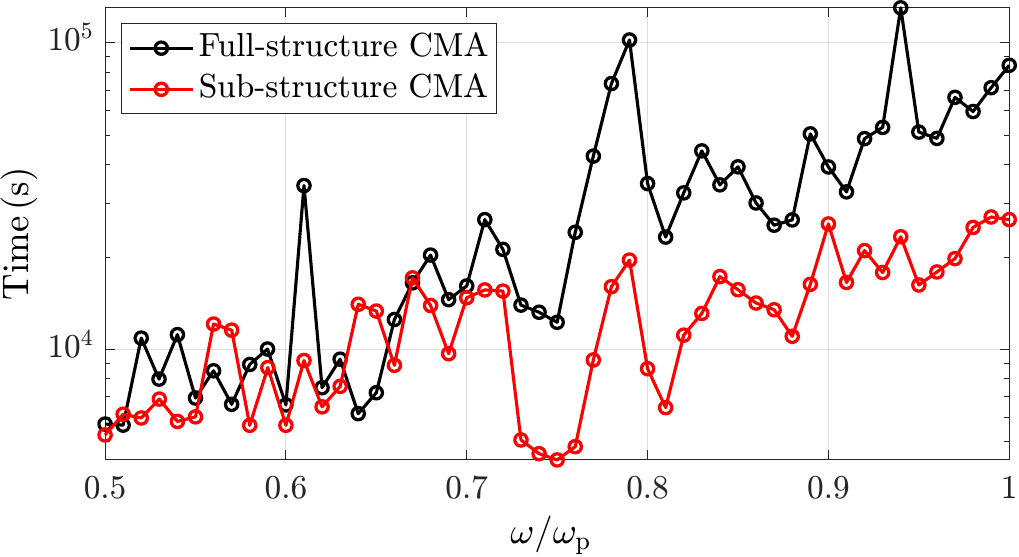}}
    \caption{(a) MS curves of the composite nanostructure obtained using the sub-structure CMA and (b) comparison of the computation times required by the full-structure and sub-structure CMA formulations.}
    \label{fig:ecs_ms_ex2_sub}
\end{figure}

The full-structure and sub-structure CMA formulations are then compared for the composite nanostructure. Because the isolated dielectric slab does not support resonances over the considered frequency range, the resonances obtained using the two formulations are expected to agree. As shown in Fig.~\ref{fig:ecs_ms_ex2}(c) and~\ref{fig:ecs_ms_ex2_sub}(a), the two formulations predict nearly identical resonance frequencies. Fig.~\ref{fig:ecs_ms_ex2_sub}(b) compares their computation times. For the sub-structure CMA, the reported time includes both the construction of the reduced matrix and the solution of the reduced GEE. The results demonstrate that, for this example, the sub-structure CMA is computationally more efficient than the full-structure CMA.

\begin{figure}[ht!]
    \centering
        \subfloat[]{\includegraphics[width=0.4\linewidth]{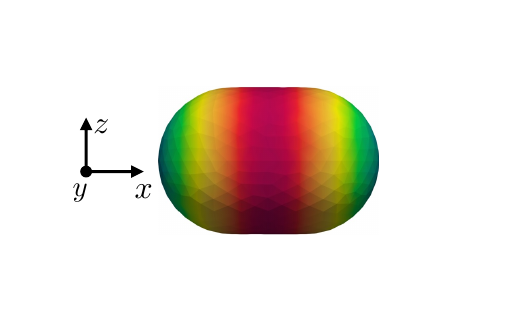}}
        \subfloat[]{\includegraphics[width=0.4\linewidth]{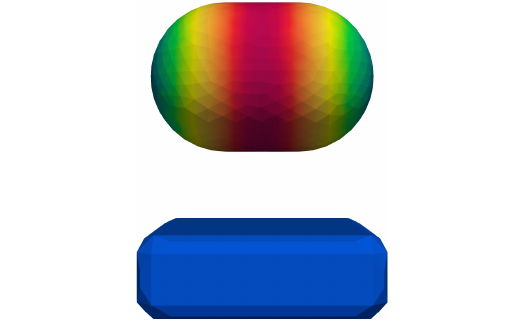}}\\
        \subfloat[]{\includegraphics[width=0.4\linewidth]{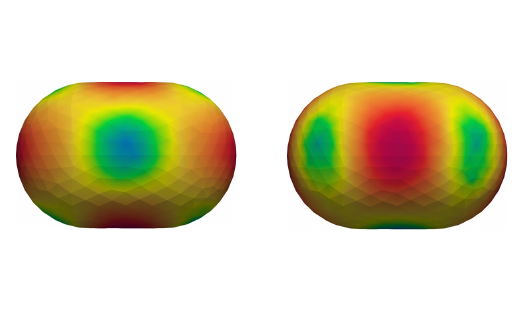}}
        \subfloat[]{\includegraphics[width=0.4\linewidth]{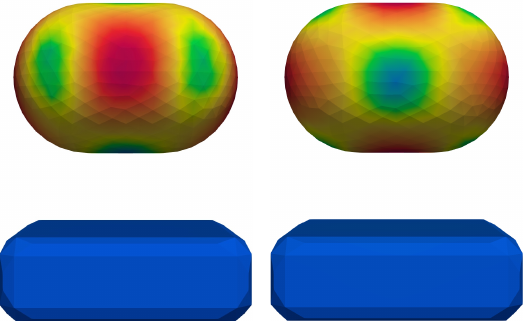}}
    \caption{Visualization of the dominant characteristic currents of the isolated metallic nanorod (left) and the composite nanostructure (right) at $\omega=0.52\,\omega_\mathrm{p}$ (a--b) and $\omega=0.70\,\omega_\mathrm{p}$ (c--d).}
    \label{fig:ex2_modes}
\end{figure}

To further examine the effect of the dielectric slab, the dominant characteristic current distributions are visualized at two resonance frequencies, $\omega=0.52\,\omega_{\mathrm{p}}$ and $\omega=0.70\,\omega_{\mathrm{p}}$, as shown in Fig.~\ref{fig:ex2_modes}. At both frequencies, the current distributions of the isolated metallic nanorod and those of the metallic region in the composite nanostructure are in very good agreement. This confirms that the dielectric slab does not significantly alter the resonant modes of the metallic nanorod. This behavior can be explained by the finite gap between the metallic and dielectric regions, together with the relatively low permittivity of the slab. The next example considers a high-contrast dielectric structure placed in direct contact with the metallic nanostructure, where a stronger modification of the metallic resonances is expected.

\subsection{Metallic Nanocube on a High-Contrast Dielectric Substrate}\label{sec:nanocube}
\begin{figure}[ht!]
\centering
  \includegraphics[width=0.6\columnwidth]{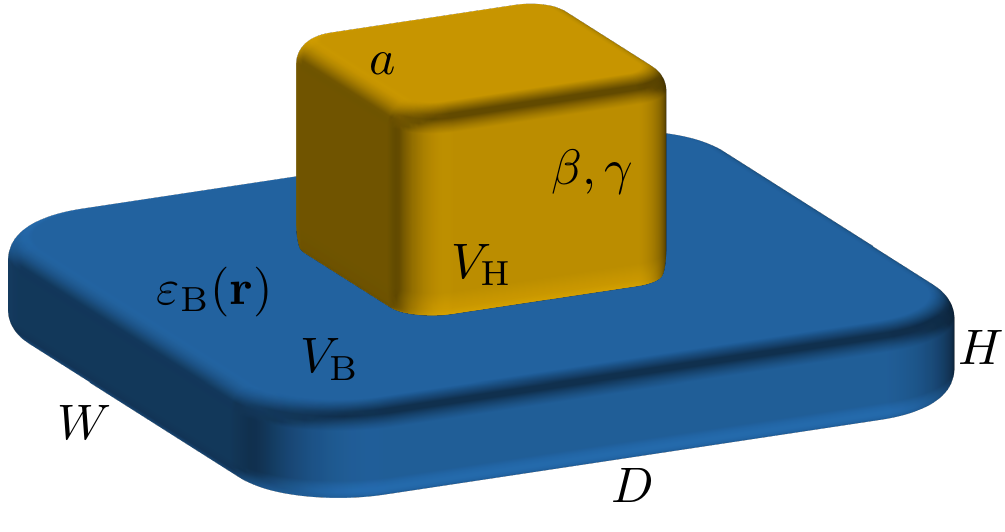}
  \caption{Composite structure comprising a gold nanocube on a high-contrast dielectric substrate.}
  \label{fig:ex3_problem}
\end{figure}

This example investigates the resonance behavior of a metallic nanocube placed on a high-contrast dielectric substrate, as shown in Fig.~\ref{fig:ex3_problem}. The metallic nanocube has a side length of $2\, \mathrm{nm}$. The dielectric substrate has dimensions of $D = 4\, \mathrm{nm}$, $W = 4\, \mathrm{nm}$, $H= 1\, \mathrm{nm}$. As in fabricated structures at nanoscale dimensions, the edges and corners of both the metallic nanocube and the dielectric substrate are rounded using cylindrical and spherical elements. This also helps to avoid geometrical singularities. The hydrodynamic parameters of the metallic nanocube are the same as those used in Section~\ref{sec:nanosphere}, and the relative permittivity of the substrate is $\varepsilon_\mathrm{B}(\mathbf{r})=7.0$. Simulations are carried out in the frequency range $[0.4,1.0]\,\omega_{\mathrm{p}}$. For the plane-wave excitation, the polarization and propagation directions are $\hat{\mathbf{p}}=\hat{\mathbf{x}}$ and $\hat{\mathbf{k}}=\hat{\mathbf{z}}$. Two cases are considered: the isolated metallic nanocube and the composite structure comprising the nanocube and the dielectric substrate. For the composite structure, the full-structure CMA formulation is used. The isolated metallic nanocube is discretized using a mesh with $31\,116$ unknowns, while the composite structure is discretized using a mesh with $N_\mathrm{B}=21\,487$ and $N_\mathrm{H}=31\,738$ unknowns. 

\begin{figure}[ht!]
    \centering   
        \subfloat[]{\includegraphics[width=0.6\columnwidth]{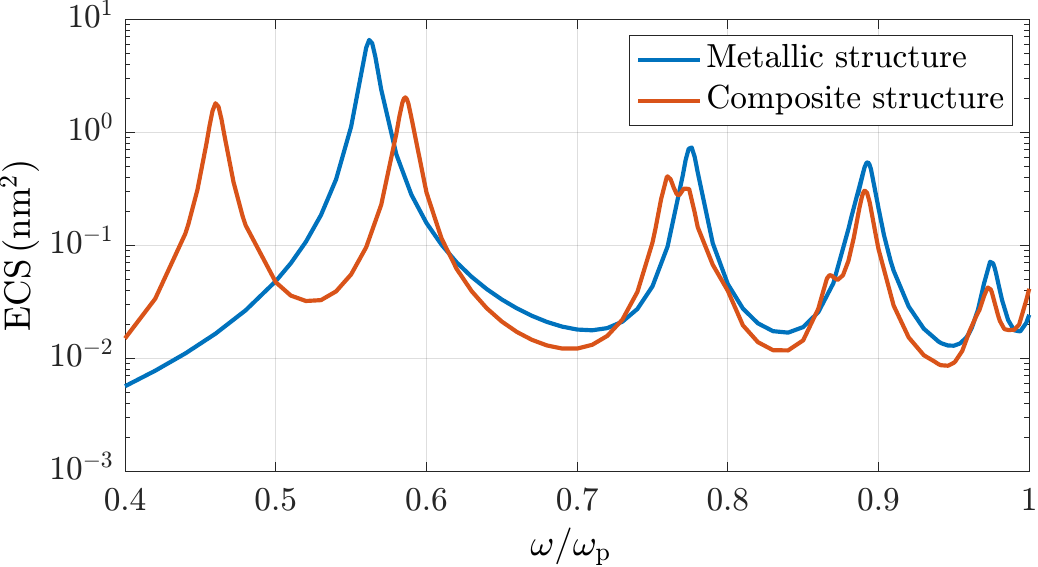}}\\
        \subfloat[]{\includegraphics[width=0.6\columnwidth]{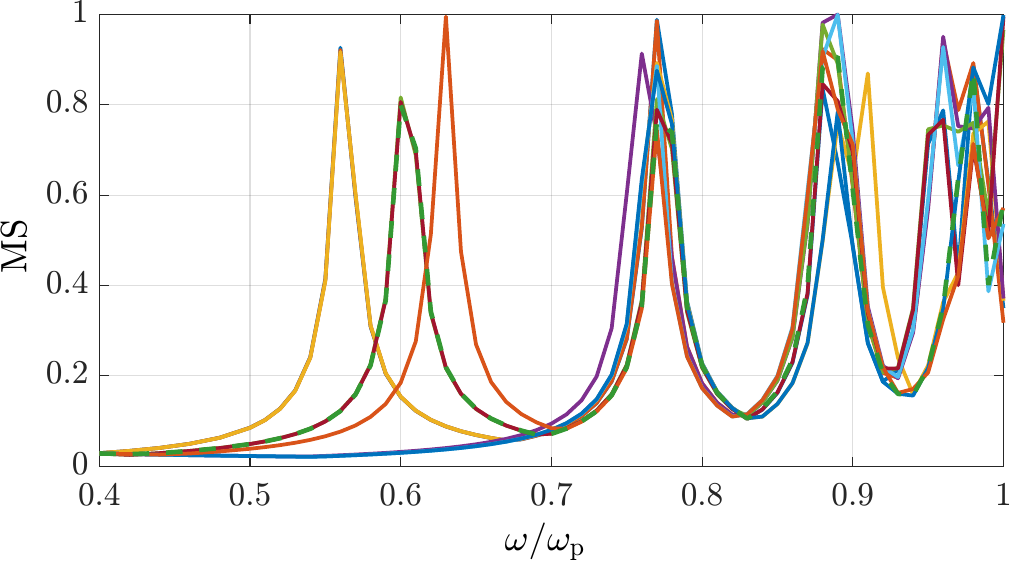}}\\
        \subfloat[]{\includegraphics[width=0.6\columnwidth]{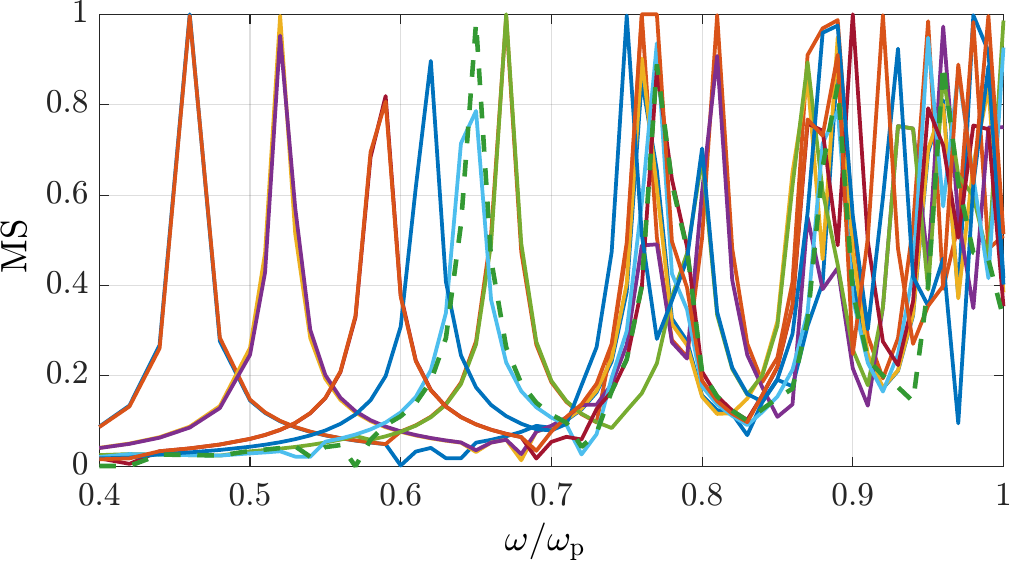}}
        \caption{(a) ECS of the isolated metallic nanocube and the composite nanostructure. MS curves of (b) the isolated metallic nanocube and (c) the composite nanostructure obtained using the full-structure CMA.}
    \label{fig:ecs_ms_ex3}
\end{figure}

Fig.~\ref{fig:ecs_ms_ex3}(a) shows the ECS of the isolated metallic nanocube and the composite nanostructure. Unlike the previous example, the resonance peaks of the isolated and composite structures differ significantly at lower frequencies, whereas the higher-frequency resonance peaks remain relatively close. This indicates that the high-contrast dielectric substrate modifies the resonance behavior of the metallic nanocube, particularly in the lower-frequency range. Fig.~\ref{fig:ecs_ms_ex3}(b) and~\ref{fig:ecs_ms_ex3}(c) show the MS curves for the isolated and composite structures, respectively. The resonance peaks observed in the MS curves agree well with the corresponding ECS peaks, confirming the validity of the CMA formulation. The comparison between the isolated and composite cases shows that the introduction of the dielectric substrate changes the modal behavior of the metallic nanocube.

\begin{figure}[ht!]
    \centering
        \subfloat[]{\includegraphics[width=0.9\columnwidth]{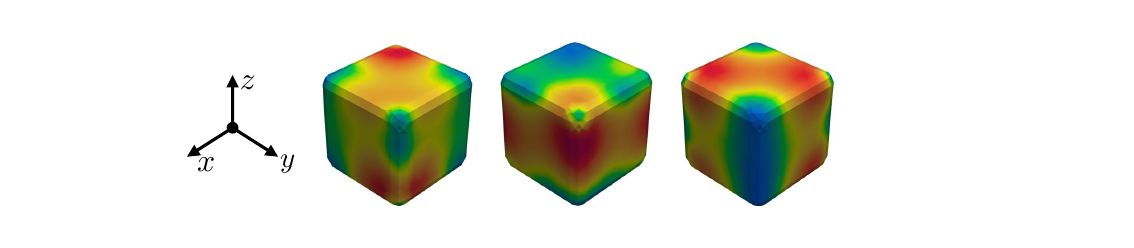}}\\
        \subfloat[]{\includegraphics[width=0.9\columnwidth]{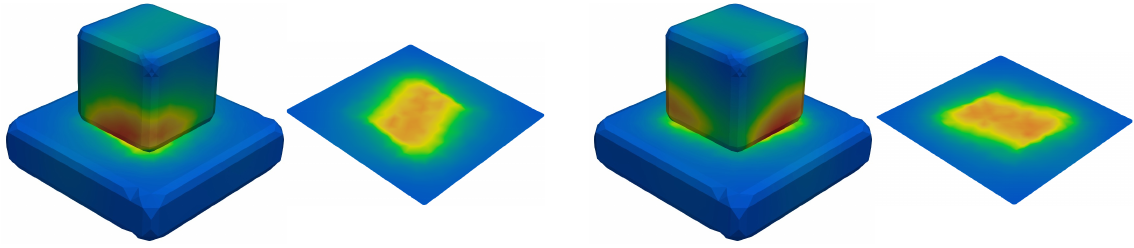}}\\
        \subfloat[]{\includegraphics[width=0.9\columnwidth]{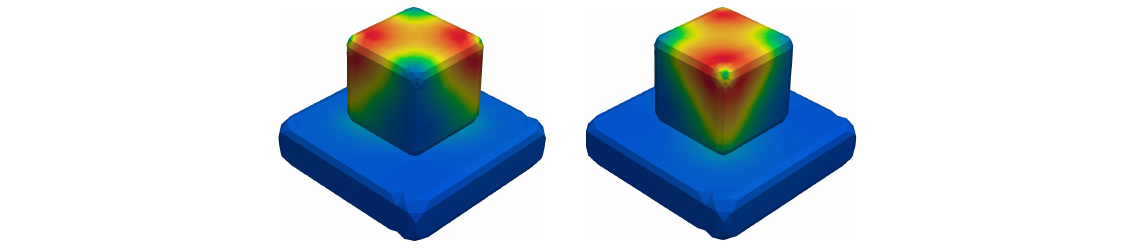}}
      \caption{Visualization of the dominant characteristic currents of (a) the isolated metallic nanocube at $\omega=0.56\,\omega_\mathrm{p}$ and the composite nanostructure at (b) $\omega=0.46\,\omega_\mathrm{p}$ and (c) $\omega=0.58\,\omega_\mathrm{p}$.}
    \label{fig:low_freq_modes_ex3}
\end{figure}

\begin{figure}[ht!]
    \centering
        \subfloat[]{\includegraphics[width=0.97\columnwidth]{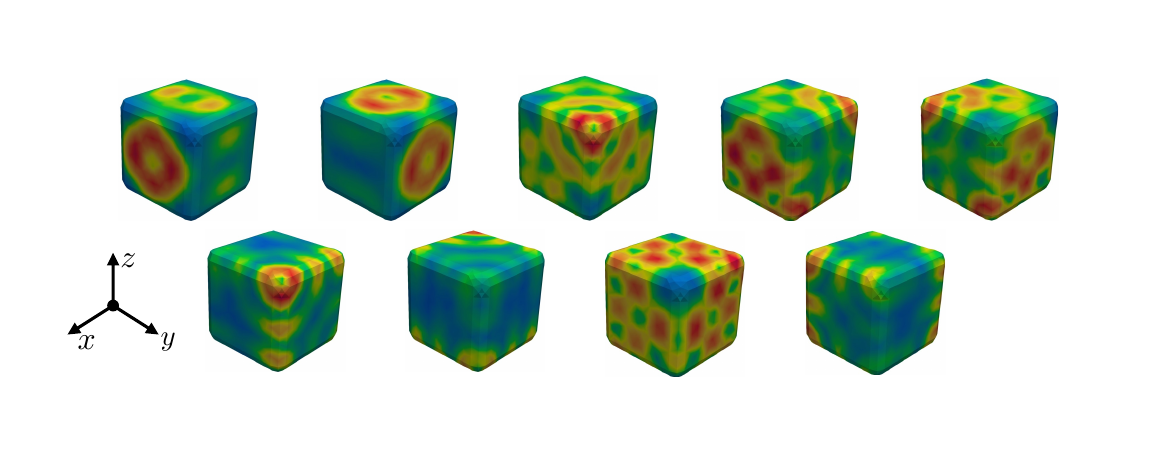}}\\
        \subfloat[]{\includegraphics[width=0.97\columnwidth]{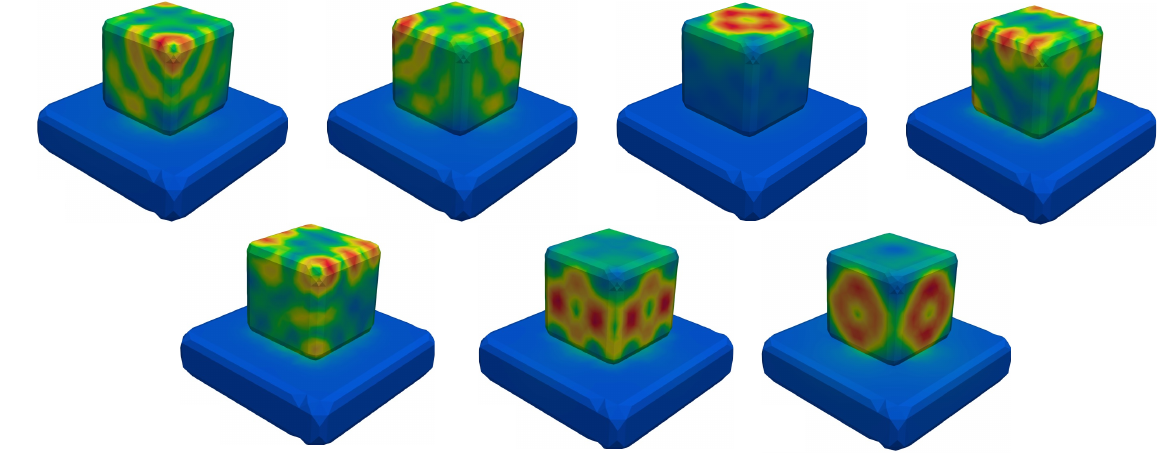}}
\caption{Visualization of the dominant characteristic currents at $\omega=0.89\,\omega_\mathrm{p}$ for (a) the isolated metallic nanocube and (b) the composite nanostructure.}
    \label{fig:high_freq_modes_ex3}
\end{figure}

To better understand the effect of the dielectric substrate, the dominant characteristic current distributions are visualized in Fig.~\ref{fig:low_freq_modes_ex3} and~\ref{fig:high_freq_modes_ex3}. Fig.~\ref{fig:low_freq_modes_ex3} compares the current distributions of the isolated metallic nanocube at $0.56\,\omega_{\mathrm{p}}$ with those of the composite structure at $0.46\,\omega_{\mathrm{p}}$ and $0.58\, \omega_{\mathrm{p}}$. This comparison shows how the dominant resonance of the isolated nanocube splits into two hybridized modes when the dielectric substrate is introduced. For the composite structure, the mode at $0.46\,\omega_{\mathrm{p}}$ exhibits stronger interaction with the substrate, as indicated by the current localization near the nanocube--substrate interface. In contrast, the mode at $0.58\,\omega_{\mathrm{p}}$ retains a current distribution closer to that of the isolated nanocube mode, although it is still perturbed by the presence of the substrate. This behavior is consistent with previous observations for metallic nanocubes on dielectric substrates, where the substrate modifies the original nanocube resonance and gives rise to hybridized modal responses~\cite{zhang2011substrate}.

Fig.~\ref{fig:high_freq_modes_ex3} shows the dominant characteristic current distributions at $0.89\,\omega_{\mathrm{p}}$ for the isolated metallic and composite structures. At this higher frequency, the current distributions of the metallic nanocube remain similar in the two cases, consistent with the ECS results showing that the higher-frequency resonance peaks are less affected by the dielectric substrate. Therefore, this example demonstrates that placing a metallic nanocube in direct contact with a high-contrast dielectric substrate can significantly modify its resonance behavior, particularly at lower frequencies.

\section{Conclusion}
\label{sec:conclusion}
This work presents full-structure and sub-structure CMA formulations for composite metallic--dielectric nanostructures, both built on the coupled VIE--HDE system that captures the nonlocal free-electron response of the metal together with the local response of the dielectric. In the full-structure formulation, the GEE is constructed from the matrix of the complete coupled system, and the resulting characteristic currents describe the response of the entire composite nanostructure. In the sub-structure formulation, the dielectric-region unknowns are eliminated through the Schur complement, yielding a reduced system whose characteristic currents are confined to the metallic region while still incorporating the full response of the dielectric region.

The numerical examples validate the formulations against the ECS and illustrate the physical behavior they capture. The influence of the dielectric medium depends on the dielectric contrast, the separation from the metal, and the contact area between the two regions. For a low-contrast dielectric separated from the metal by a finite gap, the metallic resonances are only weakly perturbed, and the full- and sub-structure formulations identify the same resonances. In this regime, the sub-structure formulation is preferred, as it is computationally more efficient. For a high-contrast substrate in direct contact with the metal, the lower-frequency resonances are strongly modified and hybridized, while the higher-frequency resonances remain largely unaffected. The main limitation of the sub-structure formulation is the cost of constructing the reduced system, which can offset its advantage for structures with large dielectric regions. 

The proposed formulations provide an effective framework for identifying and interpreting the resonances of composite metallic--dielectric nanostructures, and for quantifying how a dielectric environment reshapes the modal response of a metallic nanostructure. These capabilities are relevant to the design of plasmonic nanoantennas and nanophotonic devices, where dielectric coatings, substrates, or surrounding media are routinely used to tune resonance behavior. 

Future work will focus on reducing the computational cost of the sub-structure CMA for problems involving large dielectric regions and extending the formulation to more general material environments.

\section{Appendix}
\subsection{Physical Interpretation of Full-Structure CMA Eigenvalues}
\label{appendix:A}
Using the complex Poynting theorem, the complex power $P_{V_\mathrm{C}}$ delivered by the incident field $\mathbf{E}^\mathrm{inc}(\mathbf{r})$ to the induced bound-electron current density $\mathbf{J}_\mathrm{B}(\mathbf{r})$ in $V_\mathrm{B}$ and the free-electron current density $\mathbf{J}_\mathrm{H}(\mathbf{r})$ in $V_\mathrm{H}$ can be written as~\cite{harrington2001time}
\begin{equation}
    P_{V_\mathrm{C}}= \frac{1}{2} \int_{V_\mathrm{B}} \mathbf{E}^{\mathrm{inc}}(\mathbf{r})\cdot \mathbf{J}_\mathrm{B}^*(\mathbf{r}) \,dv
    +\frac{1}{2} \int_{V_\mathrm{H}} \mathbf{E}^{\mathrm{inc}}(\mathbf{r})\cdot \mathbf{J}_\mathrm{H}^*(\mathbf{r}) \,dv 
\label{eq:inc_power}
\end{equation}
where the superscript $*$ denotes the complex conjugate. From the complex Poynting theorem applied to the scattered electric and magnetic fields, $\mathbf{E}^{\mathrm{sca}}(\mathbf{r})$ and $\mathbf{H}^{\mathrm{sca}}(\mathbf{r})$, it follows that
\begin{equation}
    \begin{aligned}
    \frac{1}{2}\int_{V_\mathrm{B}} &\mathbf{E}^{\mathrm{sca}}(\mathbf{r})\cdot\mathbf{J}_\mathrm{B}^*(\mathbf{r})\,dv+\frac{1}{2}\int_{V_\mathrm{H}} \mathbf{E}^{\mathrm{sca}}(\mathbf{r})\cdot\mathbf{J}_\mathrm{H}^*(\mathbf{r})\,dv=\\
    & -\frac{1}{2}\oint_{\partial V_\mathrm{C}}[\mathbf{E}^{\mathrm{sca}}(\mathbf{r})\times \mathbf{H}^{\mathrm{sca}*}(\mathbf{r})]\cdot\hat{\mathbf{n}}(\mathbf{r})\,ds -\frac{\mathrm{j}\omega}{2}\int_{V_\mathrm{C}}[\mu_0|\mathbf{H}^{\mathrm{sca}}(\mathbf{r})|^2- \varepsilon_0|\mathbf{E}^{\mathrm{sca}}(\mathbf{r})|^2]\,dv.
    \end{aligned}
    \label{eq:poynting_sca}
\end{equation}
Here, $\hat{\mathbf{n}}(\mathbf{r})$ is the outward unit normal on $\partial V_\mathrm{C}$. Using~\eqref{eq:field_relation} to substitute $\mathbf{E}^\mathrm{sca}(\mathbf{r})=\mathbf{E}(\mathbf{r})-\mathbf{E}^\mathrm{inc}(\mathbf{r})$ into the left-hand side of~\eqref{eq:poynting_sca}, and using its definition in~\eqref{eq:inc_power}, $P_{V_\mathrm{C}}$ can be expressed as
\begin{equation}
    \begin{aligned}
    P_{V_\mathrm{C}} = & \frac{1}{2}\oint_{\partial V_\mathrm{C}}[\mathbf{E}^{\mathrm{sca}}(\mathbf{r})\times \mathbf{H}^{\mathrm{sca}*}(\mathbf{r})]\cdot\hat{\mathbf{n}}(\mathbf{r})\,ds +\frac{\mathrm{j}\omega}{2}\int_{V_\mathrm{C}}[\mu_0|\mathbf{H}^{\mathrm{sca}}(\mathbf{r})|^2- \varepsilon_0|\mathbf{E}^{\mathrm{sca}}(\mathbf{r})|^2]\,dv \\
    & +\frac{1}{2}\int_{V_\mathrm{B}} \mathbf{E}(\mathbf{r})\cdot\mathbf{J}_\mathrm{B}^*(\mathbf{r})\,dv +\frac{1}{2}\int_{V_\mathrm{H}} \mathbf{E}(\mathbf{r})\cdot\mathbf{J}_\mathrm{H}^*(\mathbf{r})\,dv.
    \end{aligned}
    \label{eq:power_theorem}
\end{equation}
The third term on the right-hand side of~\eqref{eq:power_theorem} can be expressed in terms of $\mathbf{E}(\mathbf{r})$ using~\eqref{eq:bound_current} as
\begin{equation}
\begin{aligned} &\int_{V_\mathrm{B}} \mathbf{E}(\mathbf{r}) \cdot\mathbf{J}^*_\mathrm{B}(\mathbf{r})\,dv =-\mathrm{j}\omega \varepsilon_0 \int_{V_\mathrm{B}}[\varepsilon_\mathrm{B}(\mathbf{r})-1]|\mathbf{E}(\mathbf{r})|^2\,dv.
\end{aligned}
\label{eq:dielectric_power} 
\end{equation}    
The last term on the right-hand side of~\eqref{eq:power_theorem} can be expressed using~\eqref{eq:hydro_current} as
\begin{equation}
\begin{aligned} 
\int_{V_\mathrm{H}}\mathbf{E}(\mathbf{r}) \cdot\mathbf{J}^*_\mathrm{H}(\mathbf{r})\,dv &=\frac{\mathrm{j}\omega}{\omega_{\mathrm{p}}^2\varepsilon_0} \int_{V_\mathrm{H}}|\mathbf{J}_\mathrm{H}(\mathbf{r})|^2\,dv- \frac{\mathrm{j}\omega\beta^2}{\omega_{\mathrm{p}}^2\varepsilon_0} \int_{V_\mathrm{H}}|\rho_\mathrm{H}(\mathbf{r})|^2\,dv \\
&+\frac{\gamma}{\omega_{\mathrm{p}}^2\varepsilon_0} \int_{V_\mathrm{H}}|\mathbf{J}_\mathrm{H}(\mathbf{r})|^2\,dv  + \frac{\beta^2}{\omega_{\mathrm{p}}^2\varepsilon_0}\int_{V_\mathrm{H}}\nabla\cdot[\rho_\mathrm{H}(\mathbf{r}) \mathbf{J}^*_\mathrm{H}(\mathbf{r})]\,dv. 
\end{aligned}
\label{eq:metal_power} 
\end{equation}    
Applying the divergence theorem to the last term in~\eqref{eq:metal_power} and enforcing the boundary condition in~\eqref{eq:bc}, the resulting surface integral vanishes. Substituting~\eqref{eq:metal_power} and~\eqref{eq:dielectric_power} into~\eqref{eq:power_theorem} and separating the resulting expression into its real and imaginary parts yields
\begin{equation}
    P_{V_\mathrm{C}} = P^{\mathrm{ext}}_{V_\mathrm{C}} + \mathrm{j} P^{\mathrm{reac}}_{V_\mathrm{C}}.
    \label{eq:power_decomp}
\end{equation}
Here, $P^{\mathrm{reac}}_{V_\mathrm{C}}$ denotes the reactive power, and the extinction power $P^{\mathrm{ext}}_{V_\mathrm{C}}$ is the sum of the scattered power $P^{\mathrm{sca}}_{V_\mathrm{C}}$ and the absorbed power $P^{\mathrm{abs}}_{V_\mathrm{H}}$~\cite{polimeridis2015}, i.e., 
\begin{equation}
    P^{\mathrm{ext}}_{V_\mathrm{C}} = P^{\mathrm{sca}}_{V_\mathrm{C}} + P^{\mathrm{abs}}_{V_\mathrm{H}}
    \label{eq:power_ext}
\end{equation}
where the absorption is confined to the metallic region $V_\mathrm{H}$.

The expressions for $P^{\mathrm{sca}}_{V_\mathrm{C}}$, $P^{\mathrm{abs}}_{V_\mathrm{H}}$, and $P^{\mathrm{reac}}_{V_\mathrm{C}}$ are given by
\begin{equation}
\begin{aligned}
    P^{\mathrm{sca}}_{V_\mathrm{C}}& = \frac{1}{2}\,\mathrm{Re}\left\{\oint_{\partial V_\mathrm{C}}[\mathbf{E}^{\mathrm{sca}}(\mathbf{r}) \times\mathbf{H}^{\mathrm{sca}*}(\mathbf{r})]\cdot\hat{\mathbf{n}}(\mathbf{r})\,ds\right\}\\
    P^{\mathrm{abs}}_{V_\mathrm{H}} &= \frac{\gamma}{2\omega_{\mathrm{p}}^2\varepsilon_0}\int_{V_\mathrm{H}}\left|\mathbf{J}_\mathrm{H}(\mathbf{r})\right|^2\,dv\\
    P^{\mathrm{reac}}_{V_\mathrm{C}}& =\frac{1}{2}\,\mathrm{Im}\left\{\oint_{\partial V_\mathrm{C}}[\mathbf{E}^{\mathrm{sca}}(\mathbf{r}) \times\mathbf{H}^{\mathrm{sca}*}(\mathbf{r})]\cdot\hat{\mathbf{n}}(\mathbf{r})\,ds\right\}\\
    &+\frac{\omega}{2}\int_{V_\mathrm{C}}[\mu_0|\mathbf{H}^{\mathrm{sca}}(\mathbf{r})|^2- \varepsilon_0|\mathbf{E}^{\mathrm{sca}}(\mathbf{r})|^2]\,dv \\
    & -\frac{\omega \varepsilon_0}{2} \int_{V_\mathrm{B}}[\varepsilon_\mathrm{B}(\mathbf{r})-1]|\mathbf{E}(\mathbf{r})|^2\,dv + \frac{\omega}{2}\left[E^\mathrm{kin} - E^\mathrm{pot}\right]
\end{aligned}
\label{eq:power_definitions}
\end{equation}
where 
\begin{equation}
\begin{aligned}
E^\mathrm{kin}& = \frac{1}{\omega_{\mathrm{p}}^2\varepsilon_0}\int_{V_\mathrm{H}}|\mathbf{J}_\mathrm{H}(\mathbf{r})|^2\,dv\\
E^\mathrm{pot}& = \frac{\beta^2}{\omega_{\mathrm{p}}^2\varepsilon_0}\int_{V_\mathrm{H}}|\rho_\mathrm{H}(\mathbf{r})|^2\,dv
\end{aligned}
\label{eq:energy_definitions}
\end{equation}
are the kinetic and potential energy contributions, respectively~\cite{forstmann1986metal}.

Substituting~\eqref{eq:current_expansion} into~\eqref{eq:inc_power}, $P_{V_\mathrm{C}}$ can be written in discretized form as
\begin{equation}
   \begin{aligned}
P_{V_\mathrm{C}} &= \frac{1}{2} \begin{bmatrix}
\bar{I}_{\mathrm{B}}^* \\
\bar{I}_{\mathrm{H}}^*
\end{bmatrix}^{\top} \begin{bmatrix}
\bar{V}^{\mathrm{inc}}_\mathrm{B}\\
\bar{V}^{\mathrm{inc}}_\mathrm{H}
\end{bmatrix} = \frac{1}{2}\bar{I}^\dagger \bar{V}^{{\mathrm{inc}}}
\end{aligned}
\label{eq:power_discrete}
\end{equation}
where the superscript $\dagger$ denotes the conjugate transpose. Using the matrix in~\eqref{eq:coupled_matrix}, $P_{V_\mathrm{C}}$ can be written as~\cite{harrington1972characteristic}
\begin{equation}
    P_{V_\mathrm{C}} =\frac{1}{2}\bar{I}^\dagger \bar{\bar{Z}} \bar{I}=\frac{1}{2}\bar{I}^\dagger  \bar{\bar{R}} \bar{I}+\mathrm{j}\frac{1}{2}\bar{I}^\dagger  \bar{\bar{X}} \bar{I}.
    \label{eq:power_impedance}
\end{equation}
It follows from~\eqref{eq:power_decomp} that the quadratic forms involving $\bar{\bar{R}}$ and $\bar{\bar{X}}$ correspond to $P^{\mathrm{ext}}_{V_\mathrm{C}}$ and $P^{\mathrm{reac}}_{V_\mathrm{C}}$, respectively~\cite{chen2015characteristic}. If the weighting matrix is chosen as $\bar{\bar{W}}=\bar{\bar{R}}$, the GEE in~\eqref{eq:gee} reduces to~\eqref{eq:gee2}. For the real characteristic currents obtained from~\eqref{eq:gee2}, the conjugate transpose reduces to the ordinary transpose. Taking the inner product of~\eqref{eq:gee2} with $\bar{I}_k$ therefore gives
\begin{equation}
    \lambda_k=\frac{\bar{I}_k^{\top}\bar{\bar{X}}\bar{I}_k}{\bar{I}_k^{\top}\bar{\bar{R}}\bar{I}_k}=\frac{P^{\mathrm{reac}}_{V_\mathrm{C},k}}{P^{\mathrm{ext}}_{V_\mathrm{C},k}}.
    \label{eq:eigenvalue}
\end{equation}
Thus, $\lambda_k$ represents the ratio between the reactive and extinction powers of the $k$th characteristic mode. Modes with $\lambda_k\approx0$ correspond to resonant modes of the composite nanostructure.

\subsection{Physical Interpretation of Sub-Structure CMA Eigenvalues}
\label{appendix:B}
In the reduced formulation, the dielectric-region unknowns are eliminated, while their influence on the metallic region is retained through the reduced matrix in~\eqref{eq:z_sub}. Consequently, the right-hand side $\bar{V}_\mathrm{sub}^\mathrm{inc}$ in~\eqref{eq:reduced_matrix} represents the effective excitation vector associated with the metallic region in the presence of the dielectric. Let $\mathbf{E}_\mathrm{sub}^{\mathrm{inc}}(\mathbf{r})$ denote the corresponding effective incident field in $V_\mathrm{H}$. The complex power associated with the metallic region is then defined as
\begin{equation}
    P_{V_\mathrm{H}}=\frac{1}{2} \int_{V_\mathrm{H}} \mathbf{E}_\mathrm{sub}^{\mathrm{inc}}(\mathbf{r}) \cdot \mathbf{J}_\mathrm{H}^*(\mathbf{r})\,dv.
    \label{eq:inc_power_metal}
\end{equation}
Similar to~\eqref{eq:power_decomp}, $P_{V_\mathrm{H}}$ can be decomposed into effective extinction and reactive components as
\begin{equation}
    P_{V_\mathrm{H}} =P^{\mathrm{ext}}_{V_\mathrm{H}} + \mathrm{j} P^{\mathrm{reac}}_{V_\mathrm{H}}.
    \label{eq:power_decomp_metal}
\end{equation}
Consistent with the effective excitation vector $\bar{V}_\mathrm{sub}^\mathrm{inc}$ in~\eqref{eq:reduced_matrix}, the discretized form of~\eqref{eq:inc_power_metal} is
\begin{equation}
P_{V_\mathrm{H}} = \frac{1}{2}\bar{I}_\mathrm{H}^\dagger \bar{V}_\mathrm{sub}^{\mathrm{inc}}.
\label{eq:power_discrete_sub}
\end{equation}
Substituting $\bar{V}_\mathrm{sub}^{\mathrm{inc}}=\bar{\bar{Z}}_\mathrm{sub}\bar{I}_\mathrm{H}$ from~\eqref{eq:reduced_matrix} into~\eqref{eq:power_discrete_sub} gives~\cite{harrington1972characteristic}
\begin{equation}
\begin{aligned}
        P_{V_\mathrm{H}}&=\frac{1}{2}\bar{I}_\mathrm{H}^\dagger \bar{\bar{Z}}_\mathrm{sub} \bar{I}_\mathrm{H}\\
        &=\frac{1}{2}\bar{I}_\mathrm{H}^\dagger  \bar{\bar{R}}_\mathrm{sub} \bar{I}_\mathrm{H}+\mathrm{j}\frac{1}{2}\bar{I}_\mathrm{H}^\dagger  \bar{\bar{X}}_\mathrm{sub} \bar{I}_\mathrm{H}.
\end{aligned}
    \label{eq:power_impedance_sub}
\end{equation}
Comparing this expression with~\eqref{eq:power_decomp_metal}, the quadratic forms associated with $\bar{\bar{R}}_\mathrm{sub}$ and $\bar{\bar{X}}_\mathrm{sub}$ correspond to $P^{\mathrm{ext}}_{V_\mathrm{H}}$ and $P^{\mathrm{reac}}_{V_\mathrm{H}}$, respectively. 

For the real characteristic currents obtained from~\eqref{eq:gee_sub}, the conjugate transpose reduces to the ordinary transpose. Taking the inner product of both sides of~\eqref{eq:gee_sub} with $\bar{I}_{\mathrm{H}_k}$ and using~\eqref{eq:power_impedance_sub}, the corresponding eigenvalue can be expressed as
\begin{equation}
    \lambda_{\mathrm{sub}_k}=\frac{\bar{I}_{\mathrm{H}_k}^{\top}\bar{\bar{X}}_\mathrm{sub}\bar{I}_{\mathrm{H}_k}}{\bar{I}_{\mathrm{H}_k}^{\top}\bar{\bar{R}}_\mathrm{sub}\bar{I}_{\mathrm{H}_k}}=\frac{P^{\mathrm{reac}}_{V_\mathrm{H},k}}{P^{\mathrm{ext}}_{V_\mathrm{H},k}}.
    \label{eq:eig_sub}
\end{equation}
Therefore, $\lambda_{\mathrm{sub}_k}$ represents the ratio between the effective reactive and extinction powers associated with the $k$th characteristic mode, $\bar{I}_{\mathrm{H}_k}$. Modes with $\lambda_{\mathrm{sub}_k}\approx0$ correspond to resonant modes of the metallic region in the presence of the dielectric region. Although the $k$th characteristic mode with coefficient vector $\bar{I}_{\mathrm{H}_k}$ is confined to $V_\mathrm{H}$, the resulting modal behavior reflects the electromagnetic interaction with the surrounding dielectric. This is because $\bar{\bar{R}}_\mathrm{sub}$ and $\bar{\bar{X}}_\mathrm{sub}$ are obtained from $\bar{\bar{Z}}_\mathrm{sub}$. Its Schur-complement term $\bar{\bar{Z}}_\mathrm{HB}\bar{\bar{Z}}_\mathrm{BB}^{-1}\bar{\bar{Z}}_\mathrm{BH}$ incorporates the full response of the dielectric region.

\subsection{Explicit Forms of the Matrix-Block and the Excitation-Vector Entries}
\label{appendix:C}
Within each tetrahedron, $\varepsilon_\mathrm{B}(\mathbf{r})$ is sampled at the centroid and held constant. For the two tetrahedra $V_n^{+}$ and $V_n^{-}$ sharing triangular face $S_n$,
\begin{equation*}
\varepsilon_{\mathrm{B},n}(\mathbf{r})=
\begin{cases}
\varepsilon_{\mathrm{B},n}^{+}=\varepsilon_{\mathrm{B}}(\mathbf{r}_{\mathrm{c}}^{+}), & \mathbf{r}\in V_{n}^{+}\\[2pt]
\varepsilon_{\mathrm{B},n}^{-}=\varepsilon_{\mathrm{B}}(\mathbf{r}_{\mathrm{c}}^{-}), & \mathbf{r}\in V_{n}^{-}
\end{cases}
\end{equation*}
where $\mathbf{r}_{\mathrm{c}}^{+}$ and $\mathbf{r}_{\mathrm{c}}^{-}$ are the centroids of $V_n^{+}$ and $V_n^{-}$.

\subsubsection{Entries of $\bar{\bar{Z}}_{\mathrm{BB}}$}
For a full testing function $\mathbf{f}_{m}^{\mathrm{B}}(\mathbf{r})$ and a full basis function $\mathbf{f}_{n}^{\mathrm{B}}(\mathbf{r})$,
\begin{equation*}
\begin{aligned}
\{\bar{\bar{Z}}_{\mathrm{BB}}\}_{mn}&=\frac{1}{\mathrm{j}\omega\varepsilon_{0}}\int_{V_{m}}\frac{\mathbf{f}_{m}^{\mathrm{B}}(\mathbf{r})\cdot\mathbf{f}_{n}^{\mathrm{B}}(\mathbf{r})}{\varepsilon_{\mathrm{B},n}(\mathbf{r})-1}\,dv+\mathrm{j}\omega\mu_{0}\int_{V_{m}}\mathbf{f}_{m}^{\mathrm{B}}(\mathbf{r})\cdot\int_{V_{n}}\mathbf{f}_{n}^{\mathrm{B}}(\mathbf{r}^{\prime})\,g_0(\mathbf{r},\mathbf{r}^{\prime})\,dv^{\prime}\,dv\\
&+\frac{1}{\mathrm{j}\omega\varepsilon_{0}}\int_{V_{m}}\nabla\cdot\mathbf{f}_{m}^{\mathrm{B}}(\mathbf{r})\int_{V_{n}}\nabla^{\prime}\cdot\mathbf{f}_{n}^{\mathrm{B}}(\mathbf{r}^{\prime})\,g_0(\mathbf{r},\mathbf{r}^{\prime})\,dv^{\prime}\,dv.
\end{aligned}
\end{equation*}
For a half testing function $\mathbf{f}_{m}^{\mathrm{B}}(\mathbf{r})$ and a full basis function $\mathbf{f}_{n}^{\mathrm{B}}(\mathbf{r})$, the testing integral is restricted to $V_m^{+}$ and the boundary term on $S_m$ appears,
\begin{equation*}
\begin{aligned}
\{\bar{\bar{Z}}_{\mathrm{BB}}\}_{mn}&=\frac{1}{\mathrm{j}\omega\varepsilon_{0}}\int_{V_{m}^{+}}\frac{\mathbf{f}_{m}^{\mathrm{B}}(\mathbf{r})\cdot\mathbf{f}_{n}^{\mathrm{B}}(\mathbf{r})}{\varepsilon_{\mathrm{B},n}(\mathbf{r})-1}\,dv+\mathrm{j}\omega\mu_{0}\int_{V_{m}^{+}}\mathbf{f}_{m}^{\mathrm{B}}(\mathbf{r})\cdot\int_{V_{n}}\mathbf{f}_{n}^{\mathrm{B}}(\mathbf{r}^{\prime})\,g_0(\mathbf{r},\mathbf{r}^{\prime})\,dv^{\prime}\,dv\\
&+\frac{1}{\mathrm{j}\omega\varepsilon_{0}}\int_{V_{m}^{+}}\nabla\cdot\mathbf{f}_{m}^{\mathrm{B}}(\mathbf{r})\int_{V_{n}}\nabla^{\prime}\cdot\mathbf{f}_{n}^{\mathrm{B}}(\mathbf{r}^{\prime})\,g_0(\mathbf{r},\mathbf{r}^{\prime})\,dv^{\prime}\,dv\\
&-\frac{1}{\mathrm{j}\omega\varepsilon_{0}}\int_{S_{m}}\hat{\mathbf{n}}_{m}(\mathbf{r})\cdot\mathbf{f}_{m}^{\mathrm{B}}(\mathbf{r})\int_{V_{n}}\nabla^{\prime}\cdot\mathbf{f}_{n}^{\mathrm{B}}(\mathbf{r}^{\prime})\,g_0(\mathbf{r},\mathbf{r}^{\prime})\,dv^{\prime}\,ds.
\end{aligned}
\end{equation*}
Here, $\hat{\mathbf{n}}_{m}(\mathbf{r})$ is the outward unit normal on $S_m$ and $\hat{\mathbf{n}}_{m}(\mathbf{r})\cdot\mathbf{f}_{m}^{\mathrm{B}}(\mathbf{r})=1$.
For a full testing function $\mathbf{f}_{m}^{\mathrm{B}}(\mathbf{r})$ and a half basis function $\mathbf{f}_{n}^{\mathrm{B}}(\mathbf{r})$, the source integral is restricted to $V_n^{+}$ and a surface-charge term on $S_n$ appears,
\begin{equation*}
\begin{aligned}
&\{\bar{\bar{Z}}_{\mathrm{BB}}\}_{mn}=\frac{1}{\mathrm{j}\omega\varepsilon_{0}(\varepsilon_{\mathrm{B},n}^{+}-1)}\int_{V_{m}}\mathbf{f}_{m}^{\mathrm{B}}(\mathbf{r})\cdot\mathbf{f}_{n}^{\mathrm{B}}(\mathbf{r})\,dv+\mathrm{j}\omega\mu_{0}\int_{V_{m}}\mathbf{f}_{m}^{\mathrm{B}}(\mathbf{r})\cdot\int_{V_{n}^{+}}\mathbf{f}_{n}^{\mathrm{B}}(\mathbf{r}^{\prime})\,g_0(\mathbf{r},\mathbf{r}^{\prime})\,dv^{\prime}\,dv\\
&+\frac{1}{\mathrm{j}\omega\varepsilon_{0}}\int_{V_{m}}\nabla\cdot\mathbf{f}_{m}^{\mathrm{B}}(\mathbf{r})\left[\int_{V_{n}^{+}}\nabla^{\prime}\cdot\mathbf{f}_{n}^{\mathrm{B}}(\mathbf{r}^{\prime})\,g_0(\mathbf{r},\mathbf{r}^{\prime})\,dv^{\prime}-\int_{S_{n}}\hat{\mathbf{n}}_{n}(\mathbf{r}^{\prime})\cdot\mathbf{f}_{n}^{\mathrm{B}}(\mathbf{r}^{\prime})\,g_0(\mathbf{r},\mathbf{r}^{\prime})\,ds^{\prime}\right]dv.
\end{aligned}
\end{equation*}
Similarly, $\hat{\mathbf{n}}_{n}(\mathbf{r}^{\prime})$ is the outward unit normal on $S_n$ and $\hat{\mathbf{n}}_{n}(\mathbf{r}^{\prime})\cdot\mathbf{f}_{n}^{\mathrm{B}}(\mathbf{r}^{\prime})=1$. For a half testing function $\mathbf{f}_{m}^{\mathrm{B}}(\mathbf{r})$ and a half basis function $\mathbf{f}_{n}^{\mathrm{B}}(\mathbf{r})$, the testing and source integrals are restricted to $V_m^{+}$ and $V_n^{+}$, respectively, and both the boundary term on $S_m$ and the surface-charge term on $S_n$ appear,
\begin{equation*}
\begin{aligned}
&\{\bar{\bar{Z}}_{\mathrm{BB}}\}_{mn}=\frac{1}{\mathrm{j}\omega\varepsilon_{0}(\varepsilon_{\mathrm{B},n}^{+}-1)}\int_{V_{m}^{+}}\mathbf{f}_{m}^{\mathrm{B}}(\mathbf{r})\cdot\mathbf{f}_{n}^{\mathrm{B}}(\mathbf{r})\,dv+\mathrm{j}\omega\mu_{0}\int_{V_{m}^{+}}\mathbf{f}_{m}^{\mathrm{B}}(\mathbf{r})\cdot\int_{V_{n}^{+}}\mathbf{f}_{n}^{\mathrm{B}}(\mathbf{r}^{\prime})\,g_0(\mathbf{r},\mathbf{r}^{\prime})\,dv^{\prime}\,dv\\
&+\frac{1}{\mathrm{j}\omega\varepsilon_{0}}\int_{V_{m}^{+}}\nabla\cdot\mathbf{f}_{m}^{\mathrm{B}}(\mathbf{r})\left[\int_{V_{n}^{+}}\nabla^{\prime}\cdot\mathbf{f}_{n}^{\mathrm{B}}(\mathbf{r}^{\prime})\,g_0(\mathbf{r},\mathbf{r}^{\prime})\,dv^{\prime}-\int_{S_{n}}\hat{\mathbf{n}}_{n}(\mathbf{r}^{\prime})\cdot\mathbf{f}_{n}^{\mathrm{B}}(\mathbf{r}^{\prime})\,g_0(\mathbf{r},\mathbf{r}^{\prime})\,ds^{\prime}\right]dv\\
&-\frac{1}{\mathrm{j}\omega\varepsilon_{0}}\int_{S_{m}}\hat{\mathbf{n}}_{m}(\mathbf{r})\cdot\mathbf{f}_{m}^{\mathrm{B}}(\mathbf{r})\left[\int_{V_{n}^{+}}\nabla^{\prime}\cdot\mathbf{f}_{n}^{\mathrm{B}}(\mathbf{r}^{\prime})\,g_0(\mathbf{r},\mathbf{r}^{\prime})\,dv^{\prime}-\int_{S_{n}}\hat{\mathbf{n}}_{n}(\mathbf{r}^{\prime})\cdot\mathbf{f}_{n}^{\mathrm{B}}(\mathbf{r}^{\prime})\,g_0(\mathbf{r},\mathbf{r}^{\prime})\,ds^{\prime}\right]ds.
\end{aligned}
\end{equation*}

\subsubsection{Entries of $\bar{\bar{Z}}_{\mathrm{BH}}$}
For a full testing function $\mathbf{f}_{m}^{\mathrm{B}}(\mathbf{r})$ and a full basis function $\mathbf{f}_{n}^{\mathrm{H}}(\mathbf{r})$,
\begin{equation*}
\begin{aligned}
\{\bar{\bar{Z}}_{\mathrm{BH}}\}_{mn}&=\mathrm{j}\omega\mu_{0}\int_{V_{m}}\mathbf{f}_{m}^{\mathrm{B}}(\mathbf{r})\cdot\int_{V_{n}}\mathbf{f}_{n}^{\mathrm{H}}(\mathbf{r}^{\prime})\,g_0(\mathbf{r},\mathbf{r}^{\prime})\,dv^{\prime}\,dv\\
&+\frac{1}{\mathrm{j}\omega\varepsilon_{0}}\int_{V_{m}}\nabla\cdot\mathbf{f}_{m}^{\mathrm{B}}(\mathbf{r})\int_{V_{n}}\nabla^{\prime}\cdot\mathbf{f}_{n}^{\mathrm{H}}(\mathbf{r}^{\prime})\,g_0(\mathbf{r},\mathbf{r}^{\prime})\,dv^{\prime}\,dv.
\end{aligned}
\end{equation*}
For a half testing function $\mathbf{f}_{m}^{\mathrm{B}}(\mathbf{r})$ and a full basis function $\mathbf{f}_{n}^{\mathrm{H}}(\mathbf{r})$, the testing integral is restricted to $V_m^{+}$ and the boundary term on $S_m$ appears,
\begin{equation*}
\begin{aligned}
\{\bar{\bar{Z}}_{\mathrm{BH}}\}_{mn}&=\mathrm{j}\omega\mu_{0}\int_{V_{m}^{+}}\mathbf{f}_{m}^{\mathrm{B}}(\mathbf{r})\cdot\int_{V_{n}}\mathbf{f}_{n}^{\mathrm{H}}(\mathbf{r}^{\prime})\,g_0(\mathbf{r},\mathbf{r}^{\prime})\,dv^{\prime}\,dv\\
&+\frac{1}{\mathrm{j}\omega\varepsilon_{0}}\int_{V_{m}^{+}}\nabla\cdot\mathbf{f}_{m}^{\mathrm{B}}(\mathbf{r})\int_{V_{n}}\nabla^{\prime}\cdot\mathbf{f}_{n}^{\mathrm{H}}(\mathbf{r}^{\prime})\,g_0(\mathbf{r},\mathbf{r}^{\prime})\,dv^{\prime}\,dv\\
&-\frac{1}{\mathrm{j}\omega\varepsilon_{0}}\int_{S_{m}}\hat{\mathbf{n}}_{m}(\mathbf{r})\cdot\mathbf{f}_{m}^{\mathrm{B}}(\mathbf{r})\int_{V_{n}}\nabla^{\prime}\cdot\mathbf{f}_{n}^{\mathrm{H}}(\mathbf{r}^{\prime})\,g_0(\mathbf{r},\mathbf{r}^{\prime})\,dv^{\prime}\,ds.
\end{aligned}
\end{equation*}

\subsubsection{Entries of $\bar{\bar{Z}}_{\mathrm{HB}}$}
For a full testing function $\mathbf{f}_{m}^{\mathrm{H}}(\mathbf{r})$ and a full basis function $\mathbf{f}_{n}^{\mathrm{B}}(\mathbf{r})$,
\begin{equation*}
\begin{aligned}
\{\bar{\bar{Z}}_{\mathrm{HB}}\}_{mn}&=\mathrm{j}\omega\mu_{0}\int_{V_{m}}\mathbf{f}_{m}^{\mathrm{H}}(\mathbf{r})\cdot\int_{V_{n}}\mathbf{f}_{n}^{\mathrm{B}}(\mathbf{r}^{\prime})\,g_0(\mathbf{r},\mathbf{r}^{\prime})\,dv^{\prime}\,dv\\
&+\frac{1}{\mathrm{j}\omega\varepsilon_{0}}\int_{V_{m}}\nabla\cdot\mathbf{f}_{m}^{\mathrm{H}}(\mathbf{r})\int_{V_{n}}\nabla^{\prime}\cdot\mathbf{f}_{n}^{\mathrm{B}}(\mathbf{r}^{\prime})\,g_0(\mathbf{r},\mathbf{r}^{\prime})\,dv^{\prime}\,dv.
\end{aligned}
\end{equation*}
For a full testing function $\mathbf{f}_{m}^{\mathrm{H}}(\mathbf{r})$ and a half basis function $\mathbf{f}_{n}^{\mathrm{B}}(\mathbf{r})$, the source integral is restricted to $V_n^{+}$ and a surface-charge term on $S_n$ appears,
\begin{equation*}
\begin{aligned}
\{\bar{\bar{Z}}_{\mathrm{HB}}\}_{mn}&=\mathrm{j}\omega\mu_{0}\int_{V_{m}}\mathbf{f}_{m}^{\mathrm{H}}(\mathbf{r})\cdot\int_{V_{n}^{+}}\mathbf{f}_{n}^{\mathrm{B}}(\mathbf{r}^{\prime})\,g_0(\mathbf{r},\mathbf{r}^{\prime})\,dv^{\prime}\,dv\\
&\!\!\!\!+\frac{1}{\mathrm{j}\omega\varepsilon_{0}}\int_{V_{m}}\nabla\cdot\mathbf{f}_{m}^{\mathrm{H}}(\mathbf{r})\left[\int_{V_{n}^{+}}\nabla^{\prime}\cdot\mathbf{f}_{n}^{\mathrm{B}}(\mathbf{r}^{\prime})\,g_0(\mathbf{r},\mathbf{r}^{\prime})\,dv^{\prime}-\int_{S_{n}}\hat{\mathbf{n}}_{n}(\mathbf{r}^{\prime})\cdot\mathbf{f}_{n}^{\mathrm{B}}(\mathbf{r}^{\prime})\,g_0(\mathbf{r},\mathbf{r}^{\prime})\,ds^{\prime}\right]dv.
\end{aligned}
\end{equation*}

\subsubsection{Entries of $\bar{\bar{Z}}_{\mathrm{HH}}$}
For a full testing function $\mathbf{f}_{m}^{\mathrm{H}}(\mathbf{r})$ and a full basis function $\mathbf{f}_{n}^{\mathrm{H}}(\mathbf{r})$,
\begin{equation*}
\begin{aligned}
\{\bar{\bar{Z}}_{\mathrm{HH}}\}_{mn}&=-\frac{\mathrm{j}\beta^{2}}{\omega\omega_{\mathrm{p}}^{2}\varepsilon_{0}}\int_{V_{m}}[\nabla\cdot\mathbf{f}_{m}^{\mathrm{H}}(\mathbf{r})][\nabla\cdot\mathbf{f}_{n}^{\mathrm{H}}(\mathbf{r})]\,dv+\frac{\mathrm{j}\omega+\gamma}{\omega_{\mathrm{p}}^{2}\varepsilon_{0}}\int_{V_{m}}\mathbf{f}_{m}^{\mathrm{H}}(\mathbf{r})\cdot\mathbf{f}_{n}^{\mathrm{H}}(\mathbf{r})\,dv\\
&+\mathrm{j}\omega\mu_{0}\int_{V_{m}}\mathbf{f}_{m}^{\mathrm{H}}(\mathbf{r})\cdot\int_{V_{n}}\mathbf{f}_{n}^{\mathrm{H}}(\mathbf{r}^{\prime})\,g_0(\mathbf{r},\mathbf{r}^{\prime})\,dv^{\prime}\,dv\\
&+\frac{1}{\mathrm{j}\omega\varepsilon_{0}}\int_{V_{m}}\nabla\cdot\mathbf{f}_{m}^{\mathrm{H}}(\mathbf{r})\int_{V_{n}}\nabla^{\prime}\cdot\mathbf{f}_{n}^{\mathrm{H}}(\mathbf{r}^{\prime})\,g_0(\mathbf{r},\mathbf{r}^{\prime})\,dv^{\prime}\,dv.
\end{aligned}
\end{equation*}

\subsubsection{Entries of $\bar{V}^{\mathrm{inc}}$}
For a full testing function $\mathbf{f}_{m}^{\mathrm{B}}(\mathbf{r})$,
\begin{equation*}
\{\bar{V}^{\mathrm{inc}}_{\mathrm{B}}\}_{m}=\int_{V_{m}}\mathbf{f}_{m}^{\mathrm{B}}(\mathbf{r})\cdot\mathbf{E}^{\mathrm{inc}}(\mathbf{r})\,dv,
\end{equation*}
and for a half testing function $\mathbf{f}_{m}^{\mathrm{B}}(\mathbf{r})$,
\begin{equation*}
\{\bar{V}^{\mathrm{inc}}_{\mathrm{B}}\}_{m}=\int_{V_{m}^{+}}\mathbf{f}_{m}^{\mathrm{B}}(\mathbf{r})\cdot\mathbf{E}^{\mathrm{inc}}(\mathbf{r})\,dv.
\end{equation*}
For a full testing function $\mathbf{f}_{m}^{\mathrm{H}}(\mathbf{r})$,
\begin{equation*}
\{\bar{V}^{\mathrm{inc}}_{\mathrm{H}}\}_{m}=\int_{V_{m}}\mathbf{f}_{m}^{\mathrm{H}}(\mathbf{r})\cdot\mathbf{E}^{\mathrm{inc}}(\mathbf{r})\,dv.
\end{equation*}

\bibliographystyle{unsrt}
\bibliography{references_final}

\newpage\clearpage

\section*{Figures}

\newpage\clearpage

\newpage\clearpage

\newpage\clearpage

\newpage\clearpage

\newpage\clearpage

\newpage\clearpage

\newpage\clearpage

\newpage\clearpage

\newpage\clearpage

\end{document}